\documentclass[10pt,journal,compsoc,finalsubmission]{IEEEtran}

\usepackage[
    style=ieee, backend=biber, isbn=false, doi=false, eprint=false, maxcitenames=2, mincitenames=1, minbibnames=1, maxbibnames=6 ]{biblatex}

\DeclareSourcemap{
	\maps[datatype=bibtex]{
		\map{
			\step[fieldset=abstract, null]
		}
\map[overwrite]{
              \step[fieldsource=shortjournal]
              \step[fieldset=journaltitle,origfieldval]
        }
        \map[overwrite]{
            \pertype{article}
            \step[fieldsource=entrykey,match=\regexp{ludwig_cress_2021},fieldset=url,fieldvalue={https://doi.org/10.36227/techrxiv.16964857.v1}]
            \step[fieldsource=entrykey,notmatch=\regexp{ludwig_cress_2021},fieldset=url, null]
        }
\map[overwrite,foreach={journal,journaltitle,booktitle}]{
          \step[fieldsource=\regexp{$MAPLOOP}, matchi={Transactions},
          replace={Trans.}]\step[fieldsource=\regexp{$MAPLOOP}, matchi={Symposium},
          replace={Symp.}]\step[fieldsource=\regexp{$MAPLOOP}, matchi={Conference},
          replace={Conf.}]\step[fieldsource=\regexp{$MAPLOOP}, matchi={Proceedings},
          replace={Proc.}]\step[fieldsource=\regexp{$MAPLOOP}, matchi={Cryptography},
          replace={Cryptogr.}]\step[fieldsource=\regexp{$MAPLOOP}, matchi={Automation},
          replace={Automat.}]\step[fieldsource=\regexp{$MAPLOOP}, matchi={Journal},
          replace={J.}]\step[fieldsource=\regexp{$MAPLOOP}, matchi={Communications},
          replace={Commun.}]\step[fieldsource=\regexp{$MAPLOOP}, matchi={Design},
          replace={Des.}]\step[fieldsource=\regexp{$MAPLOOP}, matchi={Engineering},
          replace={Eng.}]\step[fieldsource=\regexp{$MAPLOOP}, matchi={International},
          replace={Int.}]\step[fieldsource=\regexp{$MAPLOOP}, matchi={Systems},
          replace={Syst.}]\step[fieldsource=\regexp{$MAPLOOP}, matchi={Integrated},
          replace={Integr.}]\step[fieldsource=\regexp{$MAPLOOP}, matchi={Computing},
          replace={Comput.}]\step[fieldsource=\regexp{$MAPLOOP}, matchi={Computer},
          replace={Comput.}]\step[fieldsource=\regexp{$MAPLOOP}, matchi={Security},
          replace={Secur.}]\step[fieldsource=\regexp{$MAPLOOP}, matchi={Emerging},
          replace={Emerg.}]\step[fieldsource=\regexp{$MAPLOOP}, matchi={Electronic},
          replace={Electron.}]\step[fieldsource=\regexp{$MAPLOOP}, matchi=\regexp{\bon\b},
          replace={}]\step[fieldsource=\regexp{$MAPLOOP}, matchi=\regexp{\bin\b},
          replace={}]\step[fieldsource=\regexp{$MAPLOOP}, matchi=\regexp{\ba\b},
          replace={}]\step[fieldsource=\regexp{$MAPLOOP}, matchi=\regexp{\bthe\b},
          replace={}]\step[fieldsource=\regexp{$MAPLOOP}, matchi=\regexp{\bfor\b},
          replace={}]\step[fieldsource=\regexp{$MAPLOOP}, matchi=\regexp{\bof\b},
          replace={}]\step[fieldsource=\regexp{$MAPLOOP}, matchi=\regexp{\band\b},
          replace={}]}
    }
} \AtEveryBibitem{\ifentrytype{article}{\clearfield{location}\clearlist{location}
		\clearfield{address}\clearfield{publisher}\clearlist{publisher}
		\clearfield{editor}\clearfield{extra}\clearfield{pages}\clearfield{note}\clearfield{month}\clearfield{volume}\clearfield{number}\clearfield{urlyear}\clearfield{urldate}\clearfield{day}\clearfield{language}\clearfield{langid}

	}{}
	\ifentrytype{inproceedings}{\clearfield{url}\clearfield{location}\clearlist{location}
		\clearfield{publisher}\clearlist{publisher}
		\clearfield{address}\clearfield{editor}\clearfield{extra}\clearfield{pages}\clearfield{note}\clearfield{month}\clearfield{volume}\clearfield{number}\clearlist{urldate}
		\clearfield{urlyear}\clearfield{urlmonth}\clearfield{urlday}\clearfield{urldate}\clearfield{day}\clearfield{language}\clearfield{langid}

	}{}
	\ifentrytype{inbook}{\clearfield{url}\clearfield{location}\clearlist{location}
		\clearfield{publisher}\clearlist{publisher}
		\clearfield{address}\clearfield{extra}\clearfield{pages}\clearfield{note}\clearfield{month}\clearfield{number}\clearlist{urldate}
		\clearfield{urlyear}\clearfield{urlmonth}\clearfield{urlday}\clearfield{urldate}\clearfield{day}
		\clearfield{language}
		\clearfield{langid}
	
	}{}
	\ifentrytype{book}{\clearfield{url}\clearfield{location}\clearlist{location}
		\clearfield{publisher}\clearlist{publisher}
		\clearfield{address}\clearfield{extra}\clearfield{pages}\clearfield{note}\clearfield{month}\clearfield{volume}\clearfield{number}\clearfield{urlyear}\clearfield{urldate}\clearfield{language}
		\clearfield{langid}
	
	}{}
    \ifentrytype{patent}{\clearfield{url}\clearfield{location}\clearlist{location}
		\clearfield{publisher}\clearlist{publisher}
		\clearfield{address}\clearfield{extra}\clearfield{pages}\clearfield{note}\clearfield{month}\clearfield{volume}\clearfield{number}\clearfield{urlyear}\clearfield{urldate}\clearfield{language}
		\clearfield{langid}
	
	}{}
    \ifentrytype{misc}{\clearfield{location}\clearlist{location}
		\clearfield{publisher}\clearlist{publisher}
		\clearfield{address}\clearfield{extra}\clearfield{pages}\clearfield{note}\clearfield{month}\clearfield{volume}\clearfield{number}\clearfield{language}
		\clearfield{langid}
	
	}{}
    \ifentrytype{software}{\clearfield{location}\clearlist{location}
		\clearfield{publisher}\clearlist{publisher}
		\clearfield{address}\clearfield{extra}\clearfield{pages}\clearfield{note}\clearfield{month}\clearfield{volume}\clearfield{number}\clearfield{urlyear}\clearfield{urldate}\clearfield{language}
		\clearfield{langid}
	
	}{}
}

\DeclareFieldFormat{url}{\url{#1}}
\DeclareFieldFormat{titlecase}{#1}
\DeclareFieldFormat{sentencecase}{#1}

\defbibheading{References}{\section*{\refname}\addcontentsline{toc}{section}{\refname}}

\makeatletter

\AtBeginBibliography{\vskip 0.3\baselineskip plus 0.1\baselineskip minus 0.1\baselineskip}
\makeatother
\usepackage{xpatch}

\xpatchbibdriver{online}
{\printtext[parens]{\usebibmacro{date}}}
{\iffieldundef{year}{}{\printtext[parens]{\usebibmacro{date}}}}
{}{}

\usepackage[autostyle=true]{csquotes}

\usepackage{xpatch}

\makeatletter
\patchcmd\blx@bblinput{\blx@blxinit}
                      {\blx@blxinit
                      }{}{\fail}
\makeatother

\usepackage{layouts}
\usepackage{graphicx}
\usepackage{epstopdf}
\usepackage{booktabs}

\usepackage{acro}
\DeclareAcronym{ASIC}{short = ASIC , long = Application Specific Integrated Circuit}
\DeclareAcronym{RTL}{short = RTL, long = Register Transfer Level}
\DeclareAcronym{3PIP}{short = 3PIP, long = Third-Party Intellectual Property}
\DeclareAcronym{EM}{short = EM, long = electromagnetic}
\DeclareAcronym{ATPG}{short = ATPG, long = Automatic Test Pattern Generation}
\DeclareAcronym{SATsolv}{short = SAT-solving, long = Satisfiability Solving}
\DeclareAcronym{QBFsolv}{short = QBF-solving, long = Quantified Boolean Formula Solving}
\DeclareAcronym{SMTsolv}{short = SMT-solving, long = Satisfiability Modulo Theories Solving}
\DeclareAcronym{RE}{short = RE, long = Reverse Engineering}
\DeclareAcronym{SoC}{short = SoC, long = System on Chip}
\DeclareAcronym{DFS}{short = DFS, long = Depth-First-Search}
\DeclareAcronym{MCL}{short = MCL, long = Markow Cluster}
\DeclareAcronym{PCA}{short = PCA, long = Principal Component Analysis}
\DeclareAcronym{HDL}{short = HDL, long = Hardware Description Language}
\DeclareAcronym{CDF}{short = CDF, long = Cumulative Distribution Function}
\DeclareAcronym{PPF}{short = PPF, long = Percent-Point Function}
\DeclareAcronym{FSM}{short = FSM, long = Finite State Machine}
\DeclareAcronym{PRNG}{short = PRNG, long = Pseudo-Random Number Generator}
\DeclareAcronym{IQR}{short = IQR, long = Inter-Quartile Range}
\DeclareAcronym{ALU}{short = ALU, long = Arithmetic Logic Unit}
\DeclareAcronym{IO}{short = I/O, long = Input/Output}
\DeclareAcronym{AES}{short = AES, long = Advanced Encryption Standard}
\DeclareAcronym{CDMA}{short = CDMA, long = Code‐Division Multiple Access}
\DeclareAcronym{RISC}{short = RISC, long = Reduced Instruction Set Computer}
\DeclareAcronym{API}{short = API, long = Application Programming Interface}
\DeclareAcronym{HT}{short = HT, long = Hardware Trojan}
\DeclareAcronym{OCB}{short = OCB, long = Offset Codebook}
\DeclareAcronym{OCB2}{short = \acs{OCB}~2, long = \ac{OCB}~2}
\DeclareAcronym{AST}{short = AST, long = Abstract Syntax Tree}
\DeclareAcronym{NSBM}{short = NSBM, long = Nested Stochastic Block Model}
\DeclareAcronym{NN}{short = NN, long = Nearest Neighbour}
\DeclareAcronym{10NNs}{short = 10NNs, long = ten nearest neighbours}
\DeclareAcronym{NNA}{short = NNA, long = \ac{NN} Analysis}
\DeclareAcronym{QDA}{short = QDA, long = Quadratic Discriminant Analysis}
\DeclareAcronym{PDF}{short = PDF, long = Probability Density Function}
\DeclareAcronym{ROC}{short = ROC, long = Receiver Operating Characteristic}
\DeclareAcronym{RFC}{short = RFC, long = Random-Forest Classifier}
\DeclareAcronym{GBR}{short = GBR, long = Gradient Boosted Tree Regressor}
\DeclareAcronym{ADA}{short = ADA, long = AdaBoost Classifier}
\DeclareAcronym{Se}{short = Se, long = Sensitivity}
\DeclareAcronym{Sp}{short = Sp, long = Specitivity}
\DeclareAcronym{CPM}{short = CPM, long = Constant Potts Model}
\DeclareAcronym{OSHW}{short = OSHW, long = Open Source Hardware}
\DeclareAcronym{GNN}{short = GNN, long = graph neural network}
\DeclareAcronym{PUF}{short = PUF , long = Physical Unclonable Function}
\DeclareAcronym{CRP}{short = CRP, long = Challenge-Response Pair}
\DeclareAcronym{RO}{short = RO, long = ring-oscillator}
\DeclareAcronym{RO-PUF}{short = \acs{RO}-\acs{PUF}, long = \ac{RO}-\ac{PUF}}
\DeclareAcronym{FPGA}{short = FPGA, long = Field-Programmable-Gate-Array}
\DeclareAcronym{IC}{short = IC, long = Integrated Circuit}
\DeclareAcronym{TREX}{short = TREX, short-plural = es, long = Tiny Ring-Oscillator Bit Extractor}
\DeclareAcronym{IP}{short = IP, long = Intellectual Property, long-plural-form = Intellectual Properties}
\DeclareAcronym{CMOS}{short = CMOS, long = Complementary Metal-Oxide-Semiconductor}
\DeclareAcronym{NAND}{short = NAND, long = NOT-AND-Gatter, long-plural=}
\DeclareAcronym{MSB}{short = MSB, long = Most Significant Bit}
\DeclareAcronym{VHDL}{short = VHDL, long = Very High Speed Integrated Circuit Hardware Description Language}
\DeclareAcronym{DNF}{short = DNF, long = Disjunktive Normalform}
\DeclareAcronym{LUT}{short = LUT, long = Lookup-Table}
\DeclareAcronym{UART}{short = UART, long = universal asynchronous receiver/transmitter}
\DeclareAcronym{USB}{short = USB, long = Universal Serial Bus}
\DeclareAcronym{HDF5}{short = HDF5, long = Hierarchical Data Format 5}
\DeclareAcronym{LFSR}{short = LFSR, long = Linear-Feedback Shift Register}
\DeclareAcronym{NTT}{short = NTT, long = Number-Theoretic Transform}
\DeclareAcronym{SPARC}{short = SPARC, long = Scalable Processor Architecture}
\DeclareAcronym{ISA}{short = ISA, long = Instruction Set Architecture}
\DeclareAcronym{ISR}{short = ISR, long = Interrupt Service Routine}
\DeclareAcronym{LEC}{short = LEC, long = Logic Equivalence Check}
\DeclareAcronym{I2C}{short = I2C, long = Inter-Integrated Circuit}
\DeclareAcronym{FSK}{short = FSK, long = Frequency Shift Keying}
\DeclareAcronym{FF}{short = FF, long = Flip-Flop}
\DeclareAcronym{GPIO}{short = GPIO, long = General Purpose Input/Output}
\DeclareAcronym{ELF}{short = ELF, long = Executable and Linkable Format}
\DeclareAcronym{NMI}{short = NMI, long = normalised mutual information score}
\DeclareAcronym{DoS}{short = DoS, long = Denial of Service}
\DeclareAcronym{DAG}{short = DAG, long = directed acyclic graph}
\DeclareAcronym{CEC}{short = CEC, long = Circuit Equivilance Checking}
\DeclareAcronym{SAT}{short = SAT, long = Satisfiability,}
\DeclareAcronym{QBF}{short = QBF, long = Quantified Boolean Formula}
\DeclareAcronym{SMT}{short = SMT, long = Satisfiability Modulo Theories}
\DeclareAcronym{AMI}{short = AMI, long = adjusted mutual information}
\DeclareAcronym{ARI}{short = ARI, long = adjusted Rand Index}
\DeclareAcronym{nF1}{short = nF1, long = normalized F1 score}
\DeclareAcronym{CPU}{short = CPU, long = Central Processing Unit}
\DeclareAcronym{EDA}{short = EDA, long = Electronic Design Automation}
\DeclareAcronym{DFT}{short = DfT, long = Design for Test}
\DeclareAcronym{HWRE}{short = HWRE, long = Hardware Reverse Engineering}
\DeclareAcronym{USA}{short = USA, long = United States of America}
\DeclareAcronym{NDA}{short = NDA, long = Non-Disclosure Agreement}
\DeclareAcronym{RAM}{short = RAM, long = Random Access Memory}
\DeclareAcronym{GPL}{short = GPL, long = GNU General Public License}
\DeclareAcronym{RLL}{short = RLL, long = Random Logic Locking}
\DeclareAcronym{SFLL}{short = SF-LL, long = Stripped Functionality Logic Locking}
\DeclareAcronym{SEM}{short = SEM, long = Scanning Electron Microscope}
\DeclareAcronym{DRC}{short = DRC, long = Design Rule Check}
\DeclareAcronym{ERC}{short = ERC, long = Electrical Rule Check}
\DeclareAcronym{ML}{short = ML, long = machine learning}
\DeclareAcronym{GDSII}{short = GDSII, long = Graphic Design System II}
\DeclareAcronym{PCB}{short = PCB, long = Printed Circuit Board}
\DeclareAcronym{PDK}{short = PDK, long = Process Design Kit}
\DeclareAcronym{GML}{short = GML, long = golden model library}
\DeclareAcronym{KNN}{short = KNN, long = K-Nearest Neighbour}
\DeclareAcronym{CNN}{short = CNN, long = Convolutional Neural Metwork}
\DeclareAcronym{BFS}{short = BFS, long = Breadth-First Search}
\DeclareAcronym{opcode}{short = opcode, long = operation code}
\DeclareAcronym{CRESS}{short = CRESS, long = Common Reverse Engineering Scoring System}
\DeclareAcronym{CVSS}{short = CVSS, long = Common Vulnerability Scoring System}
\DeclareAcronym{IT}{short = IT, long = Information Technology}
\DeclareAcronym{DP}{short = DP, long = Data Point}
\DeclareAcronym{CWE}{short = CWE, long = Common Weakness Enumeration}
\DeclareAcronym{CVE}{short = CVE, long = Common Vulnerabilities and Exposures}
\DeclareAcronym{NVM}{short = NVM, long = Non-Volatile Memory}

\usepackage{graphicx}
\usepackage{textcomp}
\usepackage[table]{xcolor}
\def\BibTeX{{\rm B\kern-.05em{\sc i\kern-.025em b}\kern-.08em
    T\kern-.1667em\lower.7ex\hbox{E}\kern-.125emX}}

\usepackage{epstopdf}
\epstopdfDeclareGraphicsRule{.eps}{pdf}{.pdf}{epstopdf --gsopt=-dCompatibilityLevel=1.5 #1 --outfile=\OutputFile}

\usepackage{amsmath, amssymb, mathrsfs, stackrel, relsize}
\usepackage[ruled, vlined, linesnumbered]{algorithm2e}

\usepackage{balance} 

\usepackage{tikz, etoolbox, environ, ifthen, kvsetkeys, kvoptions}
\usetikzlibrary{arrows.meta}
\usetikzlibrary{decorations.pathreplacing, decorations.markings, calc, fadings}
\usetikzlibrary{backgrounds,shapes,arrows,positioning,fit}
\usetikzlibrary{tikzmark}
\usetikzlibrary{external}
\tikzsetexternalprefix{fig/}
\tikzexternalize
\tikzexternalenable

\usepackage[many]{tcolorbox}

\makeatletter
\tcbset{
	tabulars*/.style 2 args={boxsep=\z@,top=\z@,bottom=\z@,leftupper=\z@,rightupper=\z@,
		toptitle=1mm,bottomtitle=1mm,boxrule=0.5mm,
		before upper*={\let\tcb@CT@arc@save\CT@arc@ \arrayrulecolor{tcbcolframe}#1\begin{tabular}{#2}},
			after upper*={\end{tabular}\global\let\CT@arc@\tcb@CT@arc@save}
	},
	tabularcressl/.style={
        tabulars*={\setlength{\tabcolsep}{2pt}}{@{\hspace{2pt}}l@{\hspace{2pt}}},
    },
    tabularcressr/.style={
        tabulars*={\setlength{\tabcolsep}{2pt}}{@{\hspace{2pt}}>{\cellcolor{black!15!white}}l@{\hspace{2pt}}},
    },
  shield externalize
}
\makeatother

\NewDocumentCommand{\cressattrval}{mm}{\begin{tcolorbox}[
    enhanced,
    fontupper=\small\sffamily,
    arc=1pt,
    hbox,
    tcbox raise base,
    tabulars*={\setlength{\tabcolsep}{2pt}}{@{\hspace{2pt}}l>{\cellcolor{black!15!white}}l@{\hspace{2pt}}},
    nobeforeafter,
    clip upper,
    bottom=-2pt,
]
#1&#2
\end{tcolorbox}}

\makeatletter

\def\tcb@boundaryframe@pathleft{\tcb@boundary@base{0pt}{0pt}{\tcb@width}{\tcb@height}{\tcb@arc@out@SW}{\tcb@arc@out@NW}{\tcb@arc@zpt}{\tcb@arc@zpt}}
\let\tcb@drawframe@pathleft=\tcb@drawframe@path

\def\tcb@drawwithouttitle@pathleft{\tcb@pathbase{tcb fill interior}{interior.south west}{frame.east|-interior.north}{\tcb@arc@ins@SW}{\tcb@arc@ins@NW}{\tcb@arc@zpt}{\tcb@arc@zpt}}

\def\tcb@draw@border@left{\let\tcb@border=\tcb@border@left \kvtcb@borderline }

\def\tcb@border@left#1#2#3{\tcb@border@prepare{#1}{#2}\tcb@pathboundary{draw,line width=#1,#3}{\tcb@gettikzxy{([xshift=\tcb@border@ts,yshift=\tcb@border@ts]frame.south west)}{\tcb@xa}{\tcb@ya}\tcb@gettikzxy{([yshift=-\tcb@border@ts]frame.north east)}{\tcb@xb}{\tcb@yb}\pgfpathmoveto{\pgfqpoint{\tcb@xb}{\tcb@yb}}\tcb@arc@bor@NE\pgfpathlineto{\pgfqpoint{\tcb@xa}{\tcb@yb}}\tcb@arc@bor@SE\pgfpathlineto{\pgfqpoint{\tcb@xa}{\tcb@ya}}\tcb@arc@zpt\pgfpathlineto{\pgfqpoint{\tcb@xa}{\tcb@yb}}}}

\tcbset{
frame engine/pathleft/.style={frame code=\tcb@drawframe@pathleft},
interior engine/pathleft/.style={interior code=\tcb@drawwithouttitle@pathleft},
base@left/.style={clear@spec,graphical environment=tikzpicture@tcb@hooked,geometry nodes,set@outerboundary=\tcb@boundaryframe@pathleft,shape@of@skin=first,set@extensions@preframe={\kvtcb@shadow},set@extensions@postframe={\tcb@draw@border@left\tcb@apply@underlay\tcb@apply@overlay},set@extensions@final={\tcb@apply@finish}},}

\tcb@new@skin{enhancedleft}{base@left,frame engine=pathleft,interior titled engine=pathfirst,interior engine=pathleft,segmentation engine=path,title engine=pathfirst,skin first=enhancedfirst,skin middle=enhancedmiddle,skin last=enhancedmiddle}

\NewDocumentCommand{\cressattr}{m}{\begin{tcolorbox}[
    skin=enhancedleft,
    fontupper=\small\sffamily,
    arc=1pt,
    outer arc=0.5mm+1pt,
    hbox,
    tcbox raise base,
    tabularcressl,
    nobeforeafter,
    clip upper,
    bottom=-2pt,
]
#1
\end{tcolorbox}}

\def\tcb@boundaryframe@pathright{\tcb@boundary@base{0pt}{0pt}{\tcb@width}{\tcb@height}{\tcb@arc@zpt}{\tcb@arc@zpt}{\tcb@arc@out@NE}{\tcb@arc@out@SE}}
\let\tcb@drawframe@pathright=\tcb@drawframe@path

\def\tcb@drawwithouttitle@pathright{\tcb@pathbase{tcb fill interior}{frame.west|-interior.south}{interior.north east}{\tcb@arc@zpt}{\tcb@arc@zpt}{\tcb@arc@out@NE}{\tcb@arc@out@SE}}

\def\tcb@draw@border@right{\let\tcb@border=\tcb@border@right \kvtcb@borderline }

\def\tcb@border@right#1#2#3{\tcb@border@prepare{#1}{#2}\tcb@pathboundary{draw,line width=#1,#3}{\tcb@gettikzxy{([yshift=\tcb@border@ts]frame.south west)}{\tcb@xa}{\tcb@ya}\tcb@gettikzxy{([xshift=-\tcb@border@ts,yshift=-\tcb@border@ts]frame.north east)}{\tcb@xb}{\tcb@yb}\pgfpathmoveto{\pgfqpoint{\tcb@xa}{\tcb@ya}}\tcb@arc@bor@SE\pgfpathlineto{\pgfqpoint{\tcb@xb}{\tcb@ya}}\tcb@arc@bor@NE\pgfpathlineto{\pgfqpoint{\tcb@xb}{\tcb@yb}}\tcb@arc@zpt\pgfpathlineto{\pgfqpoint{\tcb@xa}{\tcb@yb}}}}

\tcbset{
frame engine/pathright/.style={frame code=\tcb@drawframe@pathright},
interior engine/pathright/.style={interior code=\tcb@drawwithouttitle@pathright},
base@right/.style={clear@spec,graphical environment=tikzpicture@tcb@hooked,geometry nodes,set@outerboundary=\tcb@boundaryframe@pathright,shape@of@skin=first,set@extensions@preframe={\kvtcb@shadow},set@extensions@postframe={\tcb@draw@border@right\tcb@apply@underlay\tcb@apply@overlay},set@extensions@final={\tcb@apply@finish}},}

\tcb@new@skin{enhancedright}{base@right,frame engine=pathright,interior titled engine=pathfirst,interior engine=pathright,segmentation engine=path,title engine=pathfirst,skin first=enhancedfirst,skin middle=enhancedmiddle,skin last=enhancedmiddle}

\NewDocumentCommand{\cressval}{m}{\begin{tcolorbox}[
    skin=enhancedright,
    fontupper=\small\sffamily,
    arc=1pt,
    hbox,
    tcbox raise base,
    tabularcressr,
    nobeforeafter,
    clip upper,
    bottom=-2pt,
    leftrule=0pt,
]
#1
\end{tcolorbox}}
\makeatother 
\usepackage{adjustbox}

\usepackage{pgfplots}
\pgfplotsset{compat=1.15}
\usepgflibrary{decorations.markings}
\usepgfplotslibrary{units}
\usepgfplotslibrary{groupplots}
\usepgfplotslibrary{colorbrewer}

\usepackage{pgfplotstable}

\usepackage{enumitem}

\usepackage{siunitx}
\usepackage{mathtools}

\usepackage{tabularx}
\usepackage{blindtext}

\setlength{\textfloatsep}{0.6cm} 
\usepackage[american]{babel}

\usepackage{color, soul}

\usepackage[hidelinks]{hyperref}

\usepackage{caption}
\usepackage{subcaption}

\usepackage{multirow}
\usepackage{multicol}

\definecolor{TUMBlau}{RGB}{0,101,189} \definecolor{TUMBlauDunkel}{RGB}{0,82,147} \definecolor{TUMBlauHell}{RGB}{152,198,234} \definecolor{TUMBlauMittel}{RGB}{100,160,200} 

\definecolor{TUMElfenbein}{RGB}{218,215,203} \definecolor{TUMGruen}{RGB}{162,173,0} \definecolor{TUMOrange}{RGB}{227,114,34} \definecolor{TUMGrau}{gray}{0.6}

\title{CRESS: Quantifying Vulnerabilities of Attack Scenarios in Hardware Reverse Engineering}

\author{Alexander Hepp, Matthias Ludwig, Michaela Brunner, Johanna Baehr, Georg Sigl\IEEEcompsocitemizethanks{\IEEEcompsocthanksitem Alexander Hepp, Matthias Ludwig, Michaela Brunner and Georg Sigl are with the TUM School of Computation, Information and Technology, Munich, Germany.\protect\\E-mail: \{alex.hepp, matthias.ludwig, michaela.brunner, sigl\}@tum.de\IEEEcompsocthanksitem Johanna Baehr and Georg Sigl are with the Fraunhofer Institute for Applied and Integrated Security (AISEC), Munich, Germany.\protect\\E-mail: johanna.baehr@aisec.fraunhofer.de}
}

\begin{document}

\IEEEtitleabstractindextext{\begin{abstract}
The safety, security, and reliability of microelectronic systems depend on a trustworthy, secured supply chain and design flow.
Globally distributed supply chains or unintentional design weaknesses leave the door open for attacks on the hardware level.
These scenarios encompass counterfeiting, hardware trojans, or on-device attacks.
For these, hardware reverse engineering (RE) results play a pivotal role.
The ongoing publication of new RE-involved attacks motivated the development of the common RE scoring system (CRESS). The system enables a general classification of RE-involved scenarios for a common, consistent rating.

In this work, the originally qualitative system is extended to a quantitative system.
We performed an extensive interview study with experts in the field.
The interview results allowed us to derive weights that measure the severity of different RE-involved attack categories.
The weights form an equation that quantifies scenarios, resulting in the severity-indicating CRESS score.
The score enables the coherent rating of novel scenarios, renders them comparable, and supports the development of effective countermeasures. To showcase the effectiveness of the quantitative CRESS Score, six selected case studies are rated qualitatively and quantitatively. The CRESS Score proves to be significantly more expressive than the industry-standard Common Vulnerability Scoring System (CVSS).
\end{abstract}

\begin{IEEEkeywords}
hardware reverse engineering, hardware security, security framework, vulnerability quantification, threat analysis, case study, IC trust, IP theft, hardware trojans
\end{IEEEkeywords}
}
\makeatletter
\def\@IEEENORMtitlevspace{1\baselineskip}
\makeatother

\maketitle \IEEEraisesectionheading{\section{Introduction}\label{sec:introduction}}

\IEEEPARstart{A}{ttack} scenarios on hardware are increasing. These include \emph{on-device} attacks (i.e.~fault injection \cite{Benso2003} or side-channel analysis \cite{Mangard2010}) and supply chain-related scenarios (i.e.~counterfeiting \cite{Tehranipoor2015}, intellectual property (IP) theft, or \acp{HT} \cite{Bhunia2017}). Academia and industry are in the throes of new or more sophisticated attack capabilities and defense methods. Hardware has evolved as a viable attack target, beyond software, for many applications. Additionally, the extremely distributed microelectronics supply chain, born of cost pressures, has a strong division of labor, which further exacerbates the situation. Potentially untrustworthy actors are in the supply chain. It is challenging to focus on safety, security, and reliability measures during distributed design and implementation. This opens the doors to malicious activities in the form of counterfeiting or \acp{HT}. Ultimately, a tradeoff between trust, i.e.~safety, security, and reliability, on the one hand, and profitability, on the other hand, is the practice. Standardized means to establish trust are available in the form of certification procedures (e.g.~common criteria (CC) \cite{CommonCriteria}, platform security architecture (PSA) \cite{PSA2023}), norms (e.g.~SAE \cite{SAE6171}, \cite{SAE6174}, IDEA-STD-1010 \cite{IDEA1010}), vulnerability enumerations (e.g.~\ac{CVE} \cite{CVE2023}, \ac{CWE} \cite{CWE2023}), or scoring schemes (e.g.~\ac{CVSS} \cite{CVSSV3_2021}).

For most hardware attack scenarios, hardware \ac{RE} plays a significant role. This role might be direct, in the form of IP theft or counterfeiting, or indirect -- i.e. supporting -- to aid \ac{HT} insertion or simplify on-device attacks. \ac{RE} research is at an all-time high, and the multitude of novel attack scenarios makes it extremely difficult to assess them. Previously mentioned standard scoring schemes allow a rough assessment of RE-involved scenarios. Yet, rating these with sufficient granularity is impossible. This motivated the development of a \ac{RE}-specific rating scheme: the common reverse engineering scoring system (CRESS) by \textcite{Ludwig2021}. 
The framework allows a general, qualitative rating of \ac{RE}-involved attack scenarios and renders them comparable. 

\IEEEpubidadjcol
In this work, we aim to extend this framework by providing the CRESS score, i.e.~the quantification of \ac{RE}-involved attack scenarios, to enable quantitative comparability.
The outline of this paper is:
\begin{itemize}
    \item First, the published  qualitative CRESS framework \cite{Ludwig2021} is introduced. This includes a brief introduction to \ac{RE} methods, their link to hardware attacks, and an explanation of the construction of the CRESS framework.
    \item For the development of an adequate equation, interviews with experts in the field were conducted. The interview is outlined in detail regarding participants, style, and questions.  
    \item The results of the interviews are evaluated and utilized to build the equation for calculating a CRESS score. The resulting CRESS equation is introduced and explained in detail.
    \item Finally, the framework is applied and evaluated via six selected case studies: A \ac{RE}-improved laser fault injection \cite{Courbon2014}, a RISC-V cryptographic chip with HTs \cite{Hepp2021}, stealthy dopant HTs \cite{Becker2013}, IP infringement and subsequent vulnerability detection, and an attack based on reading out \ac{NVM}.
\end{itemize}

\begin{figure*}

\centering
\tikzstyle{one_step} = [rectangle, draw, rotate=90, ultra thick,
   text width=.10\textwidth, text centered]    
\tikzstyle{end} = [ellipse, draw, rotate=90, ultra thick,
   text width=.088\textwidth, text centered, inner sep=-0.05mm,]    
\tikzstyle{label} = [rectangle, draw=none,  
    text width=4.9em, text centered,  minimum height=4em]
\tikzstyle{line_label} = [rectangle, draw=none, fill=white,inner sep=8pt]
\tikzstyle{line} = [draw, -latex',-{Latex[length=2mm]}, ultra thick]

\includegraphics{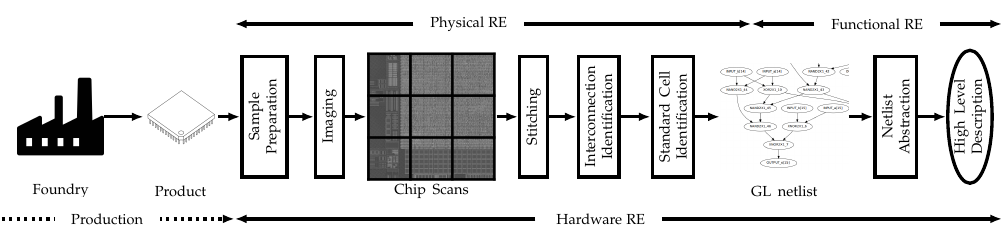}
\caption{Process flow for hardware reverse engineering of integrated circuits \cite{Ludwig2021}.}
\label{fig:REflow}
\end{figure*}
 
\section{Background}
\label{sec:background}

In the following, background on hardware reverse engineering and the architecture of the \ac{CRESS} is provided.

\subsection{A Brief RE Overview}
\label{ssec:re_overview}

Hardware \ac{RE} is the process of obtaining information of human-made devices. In the case of microelectronics, we can distinguish between system-level tear-downs, technology analyses, and layout or circuit extraction \cite{Torrance2011}. The results of the first category are component identification, bill of material creation, or simple packaging analyses. The latter two are more relevant to this work and are briefly explained below. The intermediate \ac{RE} results ($\,$\cressattr{RR}$\,$) used in CRESS are highlighted in \textit{italic}.
An overview of the \ac{RE} process is illustrated in Figure~\ref{fig:REflow}. After the design and production of microelectronic devices in semiconductor fabs or foundries, typical \ac{RE} would follow the depicted process. This process is divided into a \emph{physical} and \emph{functional} \ac{RE} part. In the remaining paper, the attributes and values comprising CRESS are highlighted as \cressattrval{Attr}{Value} (see also section \ref{ssec:state_of_the_cress}).

\subsubsection{Physical RE}
\label{sssec:physiacl_re}

Physical \ac{RE} covers both technology analyses and layout or circuit extraction. Comprehensive studies of this flow have, e.g., been conducted in \cite{Quijada2018} or \cite{Lippmann2020}. The first sub-process is the sample preparation. During \textbf{sample preparation}, the silicon die is removed from its package via a chemical or mechanical process. The die is \emph{depackaged} $\cressval{DC}$, and a \emph{dieshot} $\cressval{DS}$ can be acquired. To get sensible results in the later stages, a \emph{cross-section} $\cressval{CS}$ of the die is produced. This exposes the manufacturing parameters of the front-end or wafer-level production phase, which includes the technology node, utilized materials, and geometrical parameters. Scanning electron microscopy allows the \emph{measurement of} these \emph{technological parameters} $\cressval{TM}$. Further, if a database of technologies is available, the \emph{identification of specific technologies} $\cressval{TI}$ is possible. Additionally, this information is required for sensible delayering, which is the gradual removal of the layers of the die. Via a mixture of chemical and mechanical processes, the metallization and active layers are exposed.
For every layer, an \textbf{imaging} step is conducted. The tool for image acquisition is a high-resolution \ac{SEM}. Due to large scanning areas, a single scan is insufficient, and several (up to 1000) \emph{unstitched images} $\cressval{US}$ are acquired. In the \textbf{stitching} process, they are aligned to geometry-conserving \textit{two-dimensionally aligned mosaics} $\cressval{SL}$. Following this, adjacent layers are superimposed to create a \textit{3D stack} $\cressval{LA}$.\newline
\textbf{Interconnection identification}, synonymously called back-end-of-line extraction, is the process of segmenting conducting parts of metallization and vertical interconnect accesses via image processing (e.g.~threshold, deep learning, or other specialized algorithms) and converting them to an appropriate format, e.g.~\textit{GDSII} $\cressval{GD}$. The \textit{\textbf{identification of standard cells}} $\cressval{SI}$ is handled differently. Features of cells are extracted and clustered accordingly. For every cluster, the cells' functionality is inferred. The final stage of the physical \ac{RE} phase is the gate-level flat-netlist $\cressval{FN}$ generation, generally a gate-level graph. This graph (G) is an ordered pair G = (V, E), where V is a set of vertices (or nodes), and E is a set of edges connecting the vertices. The edges can be either directed or undirected and may have weights (values) assigned to them.\newline
Small deviations of the sequential physical \ac{RE} flow are possible. E.g.,~due to technological constraints, it might not be possible to utilize pattern recognition on the gate-level. Then, \ac{RE} must work entirely on the transistor-level netlist to perform the gate-level reconstruction via graph partitioning \cite{Putz2023}.

\subsubsection{Functional RE}
\label{sssec:functional_re}

The functional \ac{RE} process is not sequential and is an explorative process that strongly depends on the specific target of an analysis. Yet, the basis for all analyses is the product of physical \ac{RE}: the gate-level netlist or the netlist in graph representation. In the following, common analysis methods and respective results are introduced. These are in no particular order.\newline
A gate-level netlist can be \emph{partitioned} $\cressval{PN}$ by graph clustering methods \cite{Subramanyan2014}, \cite{Werner2018}. Via this approach, modules can be abstracted from the complete netlist. The identification and assignment of \emph{high-level signals} $\cressval{HS}$ allows a ``data sheet-like'' representation of signals such as input and output or the clock and reset \cite{Couch2016}. The identification of \emph{data paths} $\cressval{DI}$ \cite{Albartus2020} allows the abstraction of combined signals such as buses. From a previously partitioned netlist, \emph{functional blocks} $\cressval{PH}$ \cite{Shi2012} can be identified via structural-~or similarity-based approaches, which match these modules to a library of modules. Besides the identification of the data path, another possibility is the extraction of the \emph{control logic} $\cressval{CI}$ or finite state machine (FSM) \cite{Meade2016}, \cite{Brunner2019}. Finally, we define a \emph{partial} $\cressval{PF}$ or \emph{complete functional identification} $\cressval{CF}$ at the gate level as the last result of functional reverse engineering. This involves a perfect reconstruction allowing functional or formal verification like in the \emph{forward} design process.

The \ac{RE} result is only part of the complete CRESS architecture, which will be discussed in the following.

\subsection{State of the CRESS}
\label{ssec:state_of_the_cress}

\begin{figure}[!tbp]
    
    \definecolor{tumblue}{RGB}{0,101,189}
    \definecolor{tumdarkblue}{RGB}{0,82,147}
    \definecolor{tumdarkdarkblue}{RGB}{0,51,89}
    \definecolor{tumdarkgray}{RGB}{128,128,128}
    \definecolor{tumgray}{RGB}{204,204,204}
    \definecolor{darkgreen}{RGB}{55,157,47}
    \definecolor{darkred}{RGB}{197,47,47}
    \definecolor{tumorange}{RGB}{207,094,024}
    \definecolor{tumgreen}{RGB}{162,173,0}
    \definecolor{tumlightblue}{RGB}{100,160,200}
    \definecolor{tumlightlightblue}{RGB}{152,198,234}
    
    \makeatletter
    \pgfdeclareshape{document}{
    	\inheritsavedanchors[from=rectangle] \inheritanchorborder[from=rectangle]
    	\inheritanchor[from=rectangle]{center}
    	\inheritanchor[from=rectangle]{north}
    	\inheritanchor[from=rectangle]{south}
    	\inheritanchor[from=rectangle]{west}
    	\inheritanchor[from=rectangle]{east}
    	\inheritanchor[from=rectangle]{north east}
    	\inheritanchor[from=rectangle]{north west}
    	\inheritanchor[from=rectangle]{south east}
    	\inheritanchor[from=rectangle]{south west}
    	\backgroundpath{\southwest \pgf@xa=\pgf@x \pgf@ya=\pgf@y
    		\northeast \pgf@xb=\pgf@x \pgf@yb=\pgf@y
\pgf@xc=\pgf@xb \advance\pgf@xc by-10pt \pgf@yc=\pgf@yb \advance\pgf@yc by-10pt
\pgfpathmoveto{\pgfpoint{\pgf@xa}{\pgf@ya}}
    		\pgfpathlineto{\pgfpoint{\pgf@xa}{\pgf@yb}}
    		\pgfpathlineto{\pgfpoint{\pgf@xc}{\pgf@yb}}
    		\pgfpathlineto{\pgfpoint{\pgf@xb}{\pgf@yc}}
    		\pgfpathlineto{\pgfpoint{\pgf@xb}{\pgf@ya}}
    		\pgfpathclose
\pgfpathmoveto{\pgfpoint{\pgf@xc}{\pgf@yb}}
    		\pgfpathlineto{\pgfpoint{\pgf@xc}{\pgf@yc}}
    		\pgfpathlineto{\pgfpoint{\pgf@xb}{\pgf@yc}}
    		\pgfpathlineto{\pgfpoint{\pgf@xc}{\pgf@yc}}
    		\pgf@yd=\pgf@ya 
    		\advance\pgf@yd by5pt
    		\pgf@xc=\pgf@xa \advance\pgf@xc by5pt
    		\pgf@xd=\pgf@xb \advance\pgf@xd by-5pt
    		\pgfplotsinvokeforeach{1,...,6}{
    			\pgfpathmoveto{\pgfpoint{\pgf@xc}{\pgf@yd}}
    			\pgfpathlineto{\pgfpoint{\pgf@xd}{\pgf@yd}}
    			\advance\pgf@yd by5pt
    		}
    		\pgf@xd=\pgf@xb \advance\pgf@xd by4pt
    		\pgf@yd=\pgf@yb \advance\pgf@xd by3pt
    		\pgfpathmoveto{\pgfpoint{\pgf@xd}{\pgf@yd}}
    		\advance\pgf@xd by-10pt
    		\advance\pgf@yd by-30pt
    		\pgf@xc=\pgf@xd
    		\pgf@yc=\pgf@yd
    		\pgfpathlineto{\pgfpoint{\pgf@xd}{\pgf@yd}}
    		\advance\pgf@xd by2pt
    		\advance\pgf@yd by-1pt
    		\pgfpathlineto{\pgfpoint{\pgf@xd}{\pgf@yd}}
    		\advance\pgf@xd by10pt
    		\advance\pgf@yd by30pt
    		\pgfpathlineto{\pgfpoint{\pgf@xd}{\pgf@yd}}
    		\advance\pgf@xd by-2pt
    		\advance\pgf@yd by1pt
    		\pgfpathlineto{\pgfpoint{\pgf@xd}{\pgf@yd}}
    		\pgfpathclose
    		\pgfpathmoveto{\pgfpoint{\pgf@xc}{\pgf@yc}}
    		\advance\pgf@xc by0pt
    		\advance\pgf@yc by-4pt
    		\pgfpathlineto{\pgfpoint{\pgf@xc}{\pgf@yc}}
    		\advance\pgf@xc by2pt
    		\advance\pgf@yc by3pt
    		\pgfpathlineto{\pgfpoint{\pgf@xc}{\pgf@yc}}
    	}
    }
    \makeatother
    
    \begin{adjustbox}{width=\linewidth}
    \Large
    \includegraphics{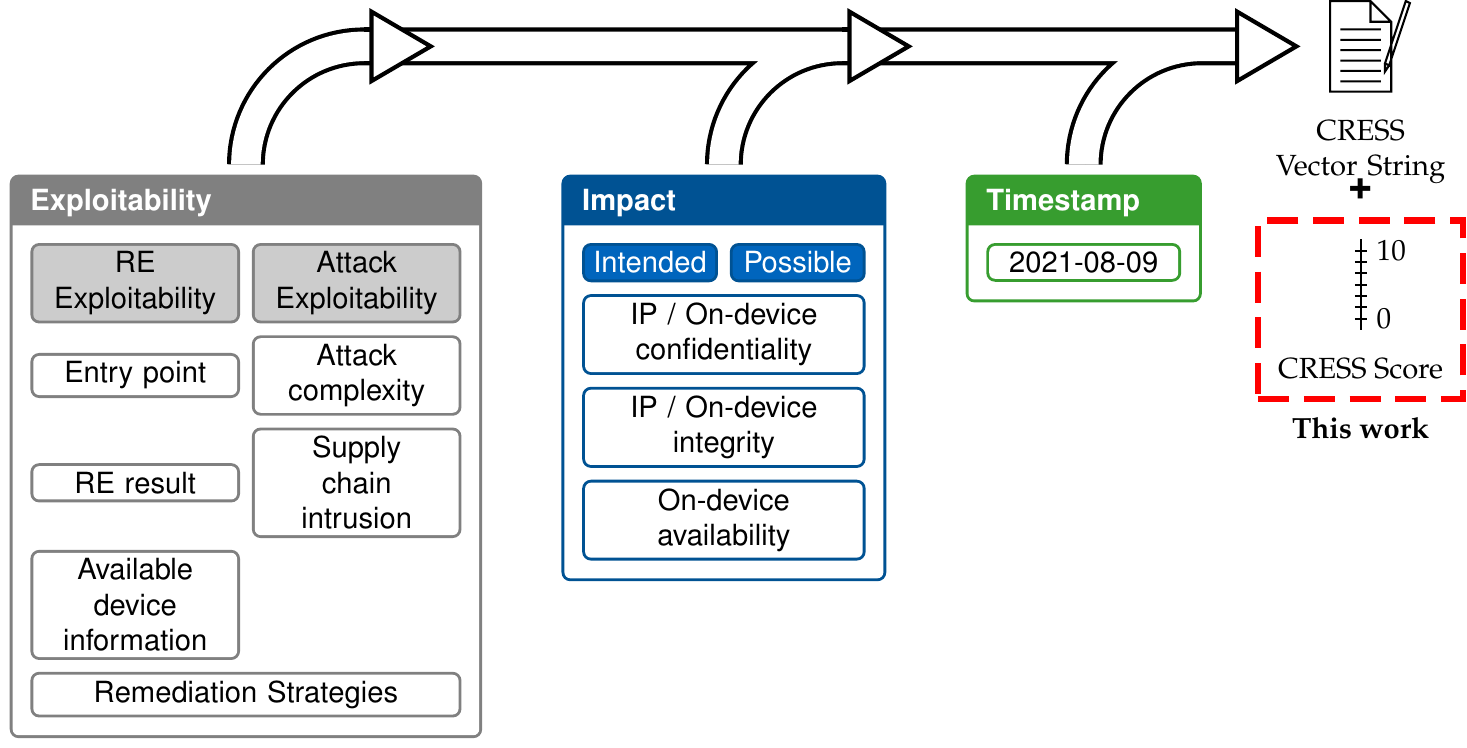}
\end{adjustbox}
\caption{Overview of the CRESS. The metrics result in a unique vector string for an individual scenario \cite{Ludwig2021}.}
	\label{fig:CRESS_BaseScore}
\end{figure}
 
The architecture of CRESS is illustrated in Figure~\ref{fig:CRESS_BaseScore}. From a high-level point-of-view, the architecture shows three main dimensions determining the CRESS vector string and score. These are the exploitability, the impact, and the timestamp. The target of the CRESS is the assignment of an attack vector by assessing the individual categories, as outlined in the following.\newline
The exploitability dimension is subdivided into two parts, of which the first is the \ac{RE} exploitability. Here, the first attribute is the entry point, i.e.~who is the attacking entity ($\,$\cressattr{EP}$\,$). The individual values are the end customer \cressval{E}, foundry \cressval{F}, physical or back-end designer \cressval{B}, front-end designer in leading \cressval{L} or partial \cressval{P} role, and third-party IP providers \cressval{I}. Next, the previously elaborated \ac{RE} result is categorized. This explains which \ac{RE} result is necessary to enable a subsequent attack. The final attribute within the \ac{RE} exploitability is the available device information ($\,$\cressattr{AI}$\,$). An attack might target a fully open source design \cressval{O}, a partially open source design \cressval{P}, or a closed source design \cressval{C}. 
Next, the attack exploitability is rated, which comprises two attributes. First, the attack complexity ($\,$\cressattr{AC}$\,$) describes the complexity of an attack that uses a given \ac{RE} result. Overall, we define five generic values: \cressval{N} none, \cressval{L} low, \cressval{M} medium, \cressval{H} high, and \cressval{X} extreme. The none category refers to an attack in which the \cressattr{RR} is directly used by the attacker, such as for IP theft, with no subsequent attack. A hobby attacker would be able to perform low-complexity hardware attacks. Next, unspecialized laboratories (medium) and specialized laboratories (high) could be involved. Finally, the extreme category describes attacks that only attack entities with unlimited resources, e.g., attacks which~state actors could perform. Via the supply chain intrusion ($\,$\cressattr{SI}$\,$), the permissions an attacking party requires for successful attack execution is defined. One option is that no intrusion is required \cressval{N}. Further, we define read-only access at selected abstraction levels \cressval{L}, read and write at a single abstraction level \cressval{M}, and read and write permission at multiple abstraction levels \cressval{H}. 
The final category included in the \ac{RE}- and the attack exploitability is remediation strategies ($\,$\cressattr{RS}$\,$). We define these as countermeasures that are effective against the given attack vector. To avoid complexity, four categories are defined. None \cressval{N} defines that no countermeasure is available. Existing countermeasures are divided into active \cressval{A**}, observing \cressval{*O*}, and logic \cressval{**L}. Active countermeasures are defined as pre-silicon countermeasures that specifically aim to thwart \ac{RE}, hardware attacks, or a combination of both. Examples are standard cell camouflaging \cite{Gomez2019} or logic locking \cite{Dupuis2019}. Observing countermeasures, mostly post-silicon methods, aim for the detection of supply chain modifications. These methods include hardware Trojan detection \cite{Hepp2021}, or avoidance and integrity or authenticity verification tools (e.g.~such as in \cite{Puschner2022} or \cite{Ludwig2023a}). Finally, the logic category defines non-technical or technical solutions aiming at attack prevention; e.g.~via access restrictions (e.g.~to photo masks or GDSII design files) or split manufacturing \cite{Vaidyanathan2014}.

The next dimension is the impact of an attack. The impact is categorized via three-letter symbols. The first letter defines whether an impact target was intended \cressattr{I**} by an described attack, or was just a possible or unintended by-product \cressattr{P**}. The second letter describes the attack target. An attack vector might aim at on-device information \cressattr{*O*}, including user data or key material. Another aim of an attack is IP \cressattr{*I*}. This includes learned information like netlists or layout data. Finally, the third letter defines the attack targets. These are \cressattr{**C} confidentiality, \cressattr{**I} integrity, and \cressattr{**A} availability. For on-device attacks, infringement of all three is possible, whereas for IP-related scenarios, only confidentiality and integrity can be targeted.

Finally, a date timestamp is added in RFC3339 format \cite{RFC3339} to account for ever-changing attack and defense mechanisms. These dimensions constitute the qualitative CRESS vector as published in \cite{Ludwig2021}. As indicated by the red dashed box in Figure~\ref{fig:CRESS_BaseScore}, the creation of the quantitative CRESS score is discussed in the following.
 \section{Methodology}
\label{sec:methodology}

The core idea of \ac{CRESS} is to render hardware vulnerabilities and attacks based on \ac{RE} comparable. The current state of \ac{CRESS} allows to compare qualitatively. To allow a quantitative comparison, the scoring results must be translated into a numeric value using an equation. This section explains how such an equation can be built and how the influence of each \ac{CRESS} attribute on the resulting score can be analyzed and decided.

\subsection{Designing an Equation for CRESS}
\label{ssec:equation_design_space}

To the best of our knowledge, there are no numerical scoring systems for reverse-engineering based hardware attacks in the literature. However, similar to \ac{CRESS}, the \ac{CVSS} \cite{CVSSV3_2021} up to version 3 uses an equation to derive the numeric score from the selected attribute values.

The \ac{CVSS} was, similar to \ac{CRESS}, built to enable a quantitative comparison of vulnerabilities, but in \ac{IT} systems. Over 3 versions, the exact composition and weighing of factors in the equations were defined more precisely, but the general composition of the equation remained untouched.

\ac{CVSS} calls its numeric score a ``Technical Severity'' \cite{CVSS_presentation}, not a risk, as it does not include a valuation of the possible damage. This severity score uses the interval $[0,10]$, where 10 stands for the maximum possible severity and 0 stands for no severity. The following details are provided by the specification document of CVSSv3.1 \cite{CVSSV3_2021}.

\ac{CVSS} builds its numeric score from two separately analyzed subscores: The exploitability score and the impact score. The exploitability explains how difficult it is to use the vulnerability or to allow the attack, while the impact scores how far an attack on the vulnerability can influence the operation of the \ac{IT} system.

Each subscore is a calculation on numeric values. Each numeric value corresponds to one \ac{CVSS} attribute. Each attribute value has an associated numeric value used in the score calculation.

While analyzing the score calculation, we noticed that the score calculation can be understood as a calculation on probabilities. The exploitability subscore is calculated by multiplying values akin to an intersection of independent probabilities. It can be understood that each exploitability value independently renders the vulnerability more or less probable, so that the exploitability probability can be calculated as the intersection of all the values.

The impact subscore is calculated by multiplying values after subtracting them from 1 and subtracting the result from 1 again. This is akin to the calculation of the union of independent probabilities by using the complements. That means that the impact of an attack using the vulnerability will cause damage related to the combination of impacts.

In the end, both subscores are combined in a weighted sum calculation (weighted arithmetic mean), akin to a mixture distribution. The final score both values the difficulty and the impact of the vulnerability with individual weights. As a result, the \ac{CVSS} equation can be understood as a weighted calculation of probabilities in the interval $[0,1]$, which is afterwards scaled to $[0,10]$.

The aim for the design of the \ac{CRESS} equation was to keep this intuitive explanation of the equation. In particular, the following aims were set for the equation design:

\begin{itemize}
    \item Explainable as a calculation on probabilities 
    \item Sensitive to changes both in Exploitability as well as Impact
    \item Produces scores across the complete interval of $[0,10]$, by scaling the final score with a factor of 10.
    \item Supports individual weights to control the influence of each attribute
\end{itemize}

A similar calculation on numeric attribute values using some combination function that allows to include weights is required for \ac{CRESS}. In general, to achieve a result in the interval $[0,1]$, such a weighted combination function must be carefully defined for given input intervals. When restricting the input intervals to $[0,1]$, the combination functions used in the \ac{CVSS} equation (multiplication, complement multiplication, arithmetic mean) achieve this property. 

While designing the equation for \ac{CRESS}, we decided to follow the intuitive pattern of \ac{CVSS} for building the equation. As a consequence, for each \ac{CRESS} attribute value an associated numeric value is required and needs to be defined. Furthermore, the influence of each attribute on the total score can and should be different, requiring a weight of influence of the attribute on the total score. To collect a maximum of expertise into the \ac{CRESS} equation, these weights and numeric values were collected with expert interviews.

It must be noted, however that the newest version 4.0 of \ac{CVSS} uses a different method for calculating the scores, that is a lookup table that encodes a total order of possible \ac{CVSS} strings and produces a vulnerability score by enumerating this total order from 0 to 10. To achieve this total order, the \ac{CVSS} 4.0 authors used various simplifications and assumptions to bring the number of possible combinations into a manageable magnitude and collected a large number of interviews for ordering this reduced list of combinations. We did not follow this approach due to the even larger number of possible combinations in \ac{CRESS}.

\subsection{Interview}
\label{ssec:interview_meth}
To derive reasonable weights and numerical values for the novel scoring equation, we take into account the expertise of various specialists in the field of hardware RE, hardware attacks, and hardware implementation. A special concern during the design and evaluation of our interviews was to receive results with minimal human bias. To address this issue, we employed several methods, including the selection of interviewees, the choice of interview style, and specific techniques during the interview to ensure honest, focused, and unbiased responses.

We aimed to interview experts with diverse backgrounds and varying areas and levels of expertise to reduce individual bias. To ensure every interviewee had an equal opportunity to respond, we scheduled individual interviews with sufficient time to cover the complete process. Furthermore, we established clear criteria for selecting interviewees, especially focusing on the expertise. More information on the interviewees can be found in section \ref{sec:results}. 

While we initially considered survey questionnaires, early tests showed that they were not ideal for effectively capturing the nuanced understanding required for the complex subject matter, leading us to opt for structured interviews instead. We conducted pilot interviews to better understand the types of biases or misunderstandings that could occur. 

Using these findings, we planned, performed, and evaluated structured interviews with volunteering experts. 
To achieve measurable and evaluable results, we developed an interview concept that was well adapted to our objective of deriving scoring equation weights and numeric values, see Section \ref{ssec:equation_design_space}. 
Additionally, the interview concept was designed to eliminate complex and confusing questions, aiming to minimize the need for long explanations and reduce the potential for misunderstandings.
Considering this, and to ensure consistency between all the interviews, we decided to use a tool called Conceptboard \cite{conceptboard} to carry out our interview in a standardized way.

\begin{figure}
     \centering
     \begin{subfigure}[b]{0.49\textwidth}
         \centering
         \includegraphics[width=\textwidth]{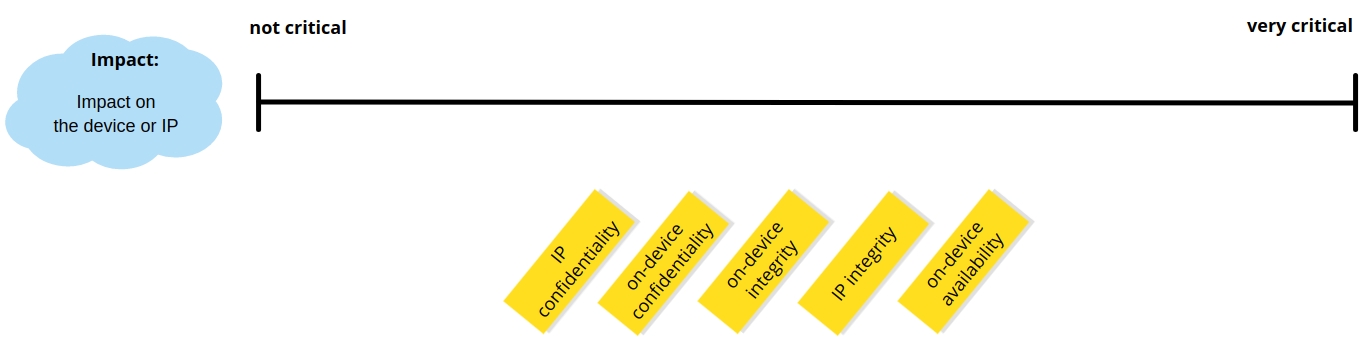}
         \caption{Exemplary question with free text box order}
         \label{fig:InterviewExamples1}
     \end{subfigure}
     \hfill
     \begin{subfigure}[b]{0.49\textwidth}
         \centering
         \includegraphics[width=\textwidth]{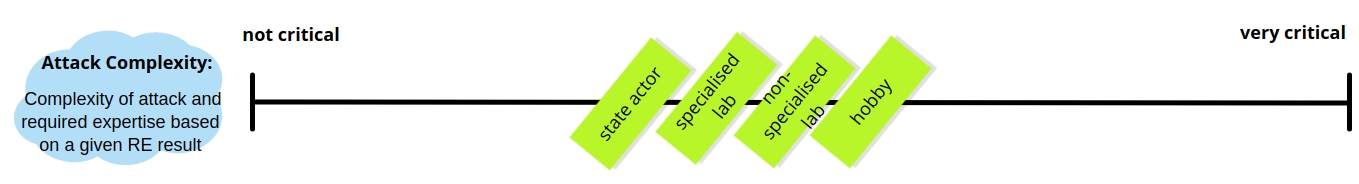}
         \caption{Exemplary question with fixed text box order}
         \label{fig:InterviewExamples2}
     \end{subfigure}
        \caption{Exemplary questions of the interactive interview using Conceptboard}
        \label{fig:InterviewExamples}
\end{figure}

Conceptboard can be used to establish an interactive interview platform.
The interviewee can answer each interview question by placing colored text boxes on a prepared scale. We chose to use scales to minimize bias, as they allow for quantified,  precise, and comparable responses, while also reducing subjectivity.
Depending on the intended granularity of the result, the order of the text boxes was either free to choose or fixed.
To rate the attack severity of the \ac{RE} result values and entry point values, and of the impact attributes, the interviewee was asked to place the text boxes on a scale from ``not critical'' to ``critical''.
The order of the values or attributes was not provided, see the example for the impact values in Figure \ref{fig:InterviewExamples1}.
To rate the attack severity of the available device information, attack complexity, and supply chain intrusion values, the interviewee also placed the text boxes on a scale from ``not critical'' to ``critical''. However, the order of the text boxes was predetermined, and it could not be altered; see the example for attack complexity in Figure \ref{fig:InterviewExamples2}.

To gain further insights into the remediation strategy attribute \cressattr{RS}, we asked the interviewees to sort various remediation strategies based on their implementation difficulty.
The interviewees could also add new remediation strategies if they felt it was necessary.

Finally, the interviewee placed a symbol for each exploitability attribute ($\,$\cressattr{EP}, $\,$\cressattr{RR}, $\,$\cressattr{AI}, $\,$\cressattr{AC}, $\,$\cressattr{SI}$\,$), a symbol for the remediation strategy \cressattr{RS}, and a symbol for the joint impact on a separate scale. In this way, the interviewee explained how strongly the respective attribute should influence the total vulnerability score.

In addition to the semi-structured interview format with predefined questions, we also included an interview guideline for the interviewer, ensuring a consistent set of questions for all interviewees to maintain uniformity in data collection. This also guarantees the best possible comparability between the different interviews. Thus, independent of the interviewer, the introduction, explanations, examples, and questions remain consistent throughout the interviews. 

The interview started with a short introduction, including information about the interviewers, the reason for the interview, a short summary of the existing CRESS framework, the interview conditions, and an overview of Conceptboard. This also ensures that the research objective of the interview is clear to the interviewee, which helps to focus their answers. 
For this purpose, we prepared an introduction slide to highlight the most important facts about CRESS, as well as a slide to test the correct functionality of Conceptboard.
During this introductory phase, we also wanted to create a comfortable environment for our participants; we utilized icebreaker questions to foster trust and encouraged honest responses. 

Next, the main part of the interview started.
In this phase, each rating question was preceded by a concise introduction, explanation and example. 
Then, the question itself, along with the action task, was presented, requiring the interviewee to assign a placement for each item as described before. We specifically phrased all interviewer text to be neutral and non-leading to avoid influencing the responses.

\begin{table*}[!htbp]
    \centering
    \caption{Overall Weight, Attribute Weight ($w_\text{attr}$) and derived Numeric Value ($V_\text{attr}$) for each Attribute Value.  Weights not directly derived from interview results are indicated by an asterisk (*).}
    \begin{tabular}{ccccccccccccc}
        \toprule
        &Attribute& Overall & Attribute & \multicolumn{9}{c}{Numeric Value ($V_\text{attr}$)} \\ 
        & & Weight & Weight ($w_\text{attr}$) & \\ 
\midrule
        \parbox[t]{0.5mm}{\multirow{14}{*}{\rotatebox[origin=c]{90}{Exploitability}}} & \multirow{2}*{Entry point \cressattr{EP}} & \multirow{14}*{$0.65$} & \multirow{2}*{$0.75$} &  \cressval{E} & \cressval{F} & \cressval{B} & \cressval{P} & \cressval{I} & \cressval{L} & & &  \\
& & & &  $0.74$ & $0.69$ & $0.48$ & $0.43$ & $0.70$ & $0.44$ &  & & \\
       
         \cline{4-13} 
        \cmidrule{4-13}
        & \multirow{4}*{RE result \cressattr{RR}} & &\multirow{4}*{$0.73$} &  \cressval{DC} & \cressval{CS} & \cressval{TM} & \cressval{TI} & \cressval{DS} & \cressval{US} & \cressval{SL} & \cressval{LA} & \cressval{GD}   \\
& & & & $0.91$ & $0.76$ & $0.79$ & $0.74$ & $0.90$ & $0.62$ & $0.51$ & $0.42$ & $0.37$    \\
\cmidrule{5-13}
         & & & &  \cressval{SI} & \cressval{FN}  & \cressval{PN} & \cressval{HS} & \cressval{PH} & \cressval{CI}& \cressval{DI} & \cressval{PF} & \cressval{CF}   \\
& & & &  $0.50$ & $0.34$ & $0.24$ & $0.35$ & $0.24$ & $0.21$ & $0.21$ & $0.11$ & $0.07$ \\
        \cline{4-13} 
        \cmidrule{4-13}
        & Available device   & & \multirow{2}*{$0.35$}&  \cressval{C} & \cressval{P} & \cressval{O} & &  & & & \\ & information \cressattr{AI} & & & $0.87$ & $0.66$ & $0.41$ & & & & & &  \\
        \cline{4-13} 
        \cmidrule{4-13}
        & Attack & & \multirow{2}*{$0.73$}& \cressval{N}* & \cressval{L} & \cressval{M} & \cressval{H} & \cressval{X}  & & &  \\
& complexity \cressattr{AC} &  & & $1.00$ & $0.93$ & $0.76$ & $0.50$ & $0.28$ & & & &    \\
        \cline{4-13} 
        \cmidrule{4-13}
        & Supply chain & & \multirow{2}*{$0.69$} & \cressval{N} & \cressval{L} & \cressval{M} & \cressval{H} & & & & &     \\
& intrusion \cressattr{SI} &  & & $0.93$ & $0.75$ & $0.50$ & $0.34$ & & &   \\
        \midrule
& Remediation & & \multirow{2}*{}& \cressval{N}* & \cressval{AOL} &  & & & & &    \\
& Strategy \cressattr{RS} & & & $1.00$ & $0.63$ &  & & & & & &    \\
\midrule
        \parbox[t]{0.5mm}{\multirow{10}{*}{\rotatebox[origin=c]{90}{Impact}}} & (I/P) IP & \multirow{10}*{$0.84$} & \multirow{2}*{$0.48$} & \cressval{H}* & \cressval{L}* & \cressval{N}* & & & & & &  \\
        & Confidentiality \cressattr{*IC} & & & $0.90$ & $0.22$ & $0.00$  & & & & & &  \\
        \cline{4-13} 
        \cmidrule{4-13}
        & \multirow{2}*{(I/P) IP Integrity \cressattr{*II}} &  & \multirow{2}*{$0.61$} & \cressval{H}* & \cressval{L}* & \cressval{N}* & & & & & &  \\
        & & & & $0.90$ & $0.22$ & $0.00$  & & & & & &  \\
        \cline{4-13} 
        \cmidrule{4-13}
        & (I/P) On-Device &  & \multirow{2}*{$0.68$} & \cressval{H}* & \cressval{L}* & \cressval{N}* & & & & & &  \\ & Confidentiality \cressattr{*OC}& & & $0.90$ & $0.22$ & $0.00$  & & & & & &  \\
        \cline{4-13} 
        \cmidrule{4-13}
        & (I/P) On-Device &  & \multirow{2}*{$0.65$} & \cressval{H}* & \cressval{L}* & \cressval{N}* & & & & & &  \\
        & Integrity \cressattr{*OI} & & & $0.90$ & $0.22$ & $0.00$  & & & & & &  \\
        \cline{4-13} 
        \cmidrule{4-13}
        & (I/P) On-Device &  & \multirow{2}*{$0.57$} & \cressval{H}* & \cressval{L}* & \cressval{N}* & & & & & &  \\
        & Availability \cressattr{*OA} & & & $0.90$ & $0.22$ & $0.00$  & & & & & &  \\
\bottomrule
    \end{tabular}
    \label{tab:CatValues}
\end{table*}

To improve the later evaluation of the results, we recorded the interview in writing with the consent of the interviewees. We consciously chose not to record the interview audio or video to encourage more open responses. The written recording allows for cross-evaluation with the answers provided in Conceptboard. Furthermore, this approach allowed for the documentation of contextual notes, which helped us interpret the data more accurately later. Consequently, each interview was conducted by a moderator and attended by a dedicated recorder

We also encouraged the interviewees to comment on their actions and think out loud while answering the questions. This made potential biases recordable and helped us understand ratings of specific values or attributes, providing us with new insights into their thought processes. 

Additionally, the final question asked the interviewee to rate their expertise in various areas related to RE and hardware security.
For this, the interviewee placed research fields, such as netlist RE, FPGA, or physical attacks, on a two-dimensional diagram with axes representing the time spent working on the topic and their level of expertise.
Additionally, the interviewee had the option to add a new research field. The final Conceptboard interview can be found at \cite{CRESSCalculator}.

\section{Results}
\label{sec:results}

This section describes how the CRESS score equation was derived from the knowledge gained in the interviews.

\subsection{Interview}
\label{ssec:interview}
We performed 21 interviews, including an interview for each of the authors. Despite the small number of interviewees, we believe their insights are significant, especially given the limited pool of experts in the field, particularly those knowledgeable in hardware reverse engineering.
To recruit volunteering specialists, we invited known experts in the field of hardware reverse engineering, attacks, or implementations in person or by email. We identified these experts through academic publications, funded projects, and industry associations related to the topic of hardware reverse engineering. We also advertised at hardware security-specific workshops. While many of the interviewees are based in Europe, experts from Asia, the Middle East, and North America were also represented.

A final review of the recordings showed that the results of two complete interviews and portions of two other interviews needed to be excluded from the final evaluation because, despite careful planning, substantial misunderstandings occurred during the interviews.
Thus, for the evaluation of the majority of the questions, we could use the data from 19 interviews. We are optimistic about the gathered results from these interviews, although we are aware that the inherent limitations of a smaller sample will be reflected in a margin of error. However, the depth and detail of the interviews, as well as special considerations taken to minimize human bias, can mitigate some of the limitations that come with this smaller sample size.

The expertise of our interviewees included several specialists in hardware RE, but also specialists in software and hardware implementations, physical/backend or digital design, failure analysis, or physical attacks. 
We asked our interviewees to assess their expertise and years of expertise across 24 categories, though they were not required to rate themselves in every category. On average, they provided ratings for 10 categories. 

\begin{figure}[t]
     \centering
     \begingroup
  \makeatletter
  \providecommand\color[2][]{\GenericError{(gnuplot) \space\space\space\@spaces}{Package color not loaded in conjunction with
      terminal option `colourtext'}{See the gnuplot documentation for explanation.}{Either use 'blacktext' in gnuplot or load the package
      color.sty in LaTeX.}\renewcommand\color[2][]{}}\providecommand\includegraphics[2][]{\GenericError{(gnuplot) \space\space\space\@spaces}{Package graphicx or graphics not loaded}{See the gnuplot documentation for explanation.}{The gnuplot epslatex terminal needs graphicx.sty or graphics.sty.}\renewcommand\includegraphics[2][]{}}\providecommand\rotatebox[2]{#2}\@ifundefined{ifGPcolor}{\newif\ifGPcolor
    \GPcolortrue
  }{}\@ifundefined{ifGPblacktext}{\newif\ifGPblacktext
    \GPblacktexttrue
  }{}\let\gplgaddtomacro\g@addto@macro
\gdef\gplbacktext{}\gdef\gplfronttext{}\makeatother
  \ifGPblacktext
\def\colorrgb#1{}\def\colorgray#1{}\else
\ifGPcolor
      \def\colorrgb#1{\color[rgb]{#1}}\def\colorgray#1{\color[gray]{#1}}\expandafter\def\csname LTw\endcsname{\color{white}}\expandafter\def\csname LTb\endcsname{\color{black}}\expandafter\def\csname LTa\endcsname{\color{black}}\expandafter\def\csname LT0\endcsname{\color[rgb]{1,0,0}}\expandafter\def\csname LT1\endcsname{\color[rgb]{0,1,0}}\expandafter\def\csname LT2\endcsname{\color[rgb]{0,0,1}}\expandafter\def\csname LT3\endcsname{\color[rgb]{1,0,1}}\expandafter\def\csname LT4\endcsname{\color[rgb]{0,1,1}}\expandafter\def\csname LT5\endcsname{\color[rgb]{1,1,0}}\expandafter\def\csname LT6\endcsname{\color[rgb]{0,0,0}}\expandafter\def\csname LT7\endcsname{\color[rgb]{1,0.3,0}}\expandafter\def\csname LT8\endcsname{\color[rgb]{0.5,0.5,0.5}}\else
\def\colorrgb#1{\color{black}}\def\colorgray#1{\color[gray]{#1}}\expandafter\def\csname LTw\endcsname{\color{white}}\expandafter\def\csname LTb\endcsname{\color{black}}\expandafter\def\csname LTa\endcsname{\color{black}}\expandafter\def\csname LT0\endcsname{\color{black}}\expandafter\def\csname LT1\endcsname{\color{black}}\expandafter\def\csname LT2\endcsname{\color{black}}\expandafter\def\csname LT3\endcsname{\color{black}}\expandafter\def\csname LT4\endcsname{\color{black}}\expandafter\def\csname LT5\endcsname{\color{black}}\expandafter\def\csname LT6\endcsname{\color{black}}\expandafter\def\csname LT7\endcsname{\color{black}}\expandafter\def\csname LT8\endcsname{\color{black}}\fi
  \fi
    \setlength{\unitlength}{0.0500bp}\ifx\gptboxheight\undefined \newlength{\gptboxheight}\newlength{\gptboxwidth}\newsavebox{\gptboxtext}\fi \setlength{\fboxrule}{0.5pt}\setlength{\fboxsep}{1pt}\definecolor{tbcol}{rgb}{1,1,1}\begin{picture}(5044.00,3684.00)\gplgaddtomacro\gplbacktext{\csname LTb\endcsname \put(1474,1083){\makebox(0,0)[r]{\strut{}1 month}}\put(1474,3304){\makebox(0,0)[r]{\strut{}10 years}}\put(1723,704){\makebox(0,0){\strut{}Novice}}\put(4521,704){\makebox(0,0){\strut{}Expert}}}\gplgaddtomacro\gplfronttext{\csname LTb\endcsname \put(539,2193){\rotatebox{-270}{\makebox(0,0){\strut{}Years of Experience}}}\put(3122,374){\makebox(0,0){\strut{}Level of Expertise }}\put(3122,154){\makebox(0,0){\strut{}}}}\gplbacktext
    \put(0,0){\includegraphics[width={252.20bp},height={184.20bp}]{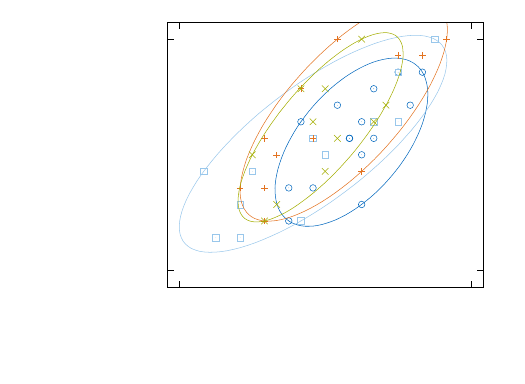}}\gplfronttext
  \end{picture}\endgroup
      \caption{Self-rated Interviewee expertise and experience for \textcolor[RGB]{0, 101, 189}{netlist RE}, \textcolor[RGB]{152, 198, 234}{physical RE}, \textcolor[RGB]{227, 114, 34}{digital design}, and \textcolor[RGB]{162, 173, 0}{hardware attacks}.}
    \label{fig:expertise}
\end{figure}

The responses for four of the more commonly chosen categories are illustrated in Figure.  \ref{fig:expertise}. 
For example, many of the interviewees rated themselves with a high level of expertise and long experience in netlist reverse engineering, whereas for physical reverse engineering the ratings were more spread out.  

\subsection{Interview Evaluation}
\label{ssec:interview_results}
For the final evaluation, the Conceptboard slides with the placed text boxes were translated into a tabular format.
In each interview and question, for each text box, a value in  $[0,1]$ is assigned, which describes the relative horizontal position on the scale (see Figure~\ref{fig:InterviewExamples}).
In the following, we will refer to this value as \ac{DP}.~These \acp{DP} are then used to derive the weights and numeric values for the scoring equation. All weights and numeric values are shown in Table \ref{tab:CatValues}. We  also provide the anonymized raw data for each interviewee \cite{CRESSCalculator}.

\begin{figure*}[!htbp]
     \centering
     \begingroup
  \makeatletter
  \providecommand\color[2][]{\GenericError{(gnuplot) \space\space\space\@spaces}{Package color not loaded in conjunction with
      terminal option `colourtext'}{See the gnuplot documentation for explanation.}{Either use 'blacktext' in gnuplot or load the package
      color.sty in LaTeX.}\renewcommand\color[2][]{}}\providecommand\includegraphics[2][]{\GenericError{(gnuplot) \space\space\space\@spaces}{Package graphicx or graphics not loaded}{See the gnuplot documentation for explanation.}{The gnuplot epslatex terminal needs graphicx.sty or graphics.sty.}\renewcommand\includegraphics[2][]{}}\providecommand\rotatebox[2]{#2}\@ifundefined{ifGPcolor}{\newif\ifGPcolor
    \GPcolortrue
  }{}\@ifundefined{ifGPblacktext}{\newif\ifGPblacktext
    \GPblacktexttrue
  }{}\let\gplgaddtomacro\g@addto@macro
\gdef\gplbacktext{}\gdef\gplfronttext{}\makeatother
  \ifGPblacktext
\def\colorrgb#1{}\def\colorgray#1{}\else
\ifGPcolor
      \def\colorrgb#1{\color[rgb]{#1}}\def\colorgray#1{\color[gray]{#1}}\expandafter\def\csname LTw\endcsname{\color{white}}\expandafter\def\csname LTb\endcsname{\color{black}}\expandafter\def\csname LTa\endcsname{\color{black}}\expandafter\def\csname LT0\endcsname{\color[rgb]{1,0,0}}\expandafter\def\csname LT1\endcsname{\color[rgb]{0,1,0}}\expandafter\def\csname LT2\endcsname{\color[rgb]{0,0,1}}\expandafter\def\csname LT3\endcsname{\color[rgb]{1,0,1}}\expandafter\def\csname LT4\endcsname{\color[rgb]{0,1,1}}\expandafter\def\csname LT5\endcsname{\color[rgb]{1,1,0}}\expandafter\def\csname LT6\endcsname{\color[rgb]{0,0,0}}\expandafter\def\csname LT7\endcsname{\color[rgb]{1,0.3,0}}\expandafter\def\csname LT8\endcsname{\color[rgb]{0.5,0.5,0.5}}\else
\def\colorrgb#1{\color{black}}\def\colorgray#1{\color[gray]{#1}}\expandafter\def\csname LTw\endcsname{\color{white}}\expandafter\def\csname LTb\endcsname{\color{black}}\expandafter\def\csname LTa\endcsname{\color{black}}\expandafter\def\csname LT0\endcsname{\color{black}}\expandafter\def\csname LT1\endcsname{\color{black}}\expandafter\def\csname LT2\endcsname{\color{black}}\expandafter\def\csname LT3\endcsname{\color{black}}\expandafter\def\csname LT4\endcsname{\color{black}}\expandafter\def\csname LT5\endcsname{\color{black}}\expandafter\def\csname LT6\endcsname{\color{black}}\expandafter\def\csname LT7\endcsname{\color{black}}\expandafter\def\csname LT8\endcsname{\color{black}}\fi
  \fi
    \setlength{\unitlength}{0.0500bp}\ifx\gptboxheight\undefined \newlength{\gptboxheight}\newlength{\gptboxwidth}\newsavebox{\gptboxtext}\fi \setlength{\fboxrule}{0.5pt}\setlength{\fboxsep}{1pt}\definecolor{tbcol}{rgb}{1,1,1}\begin{picture}(10282.00,4534.00)\gplgaddtomacro\gplbacktext{\csname LTb\endcsname \put(330,2707){\makebox(0,0)[r]{\strut{}$0$}}\put(330,3495){\makebox(0,0)[r]{\strut{}$1$}}\put(751,2487){\makebox(0,0){\strut{}E}}\put(1039,2487){\makebox(0,0){\strut{}F}}\put(1328,2487){\makebox(0,0){\strut{}B}}\put(1616,2487){\makebox(0,0){\strut{}P}}\put(1905,2487){\makebox(0,0){\strut{}I}}\put(2193,2487){\makebox(0,0){\strut{}L}}}\gplgaddtomacro\gplfronttext{\csname LTb\endcsname \put(1472,4203){\makebox(0,0){\strut{}Entry Point }}\put(1472,3983){\makebox(0,0){\strut{}}}}\gplgaddtomacro\gplbacktext{\csname LTb\endcsname \put(3208,2707){\makebox(0,0)[r]{\strut{}$0$}}\put(3208,3495){\makebox(0,0)[r]{\strut{}$1$}}\put(3684,2487){\makebox(0,0){\strut{}DC}}\put(4029,2487){\makebox(0,0){\strut{}CS}}\put(4373,2487){\makebox(0,0){\strut{}TM}}\put(4718,2487){\makebox(0,0){\strut{}TI}}\put(5062,2487){\makebox(0,0){\strut{}DS}}\put(5407,2487){\makebox(0,0){\strut{}US}}\put(5751,2487){\makebox(0,0){\strut{}SL}}\put(6096,2487){\makebox(0,0){\strut{}LA}}\put(6440,2487){\makebox(0,0){\strut{}GD}}\put(6785,2487){\makebox(0,0){\strut{}SI}}\put(7129,2487){\makebox(0,0){\strut{}FN}}\put(7474,2487){\makebox(0,0){\strut{}PN}}\put(7818,2487){\makebox(0,0){\strut{}HS}}\put(8163,2487){\makebox(0,0){\strut{}PH}}\put(8507,2487){\makebox(0,0){\strut{}CI}}\put(8852,2487){\makebox(0,0){\strut{}DI}}\put(9196,2487){\makebox(0,0){\strut{}PF}}\put(9541,2487){\makebox(0,0){\strut{}CF}}}\gplgaddtomacro\gplfronttext{\csname LTb\endcsname \put(6612,4203){\makebox(0,0){\strut{}RE Result }}\put(6612,3983){\makebox(0,0){\strut{} }}}\gplgaddtomacro\gplbacktext{\csname LTb\endcsname \put(330,440){\makebox(0,0)[r]{\strut{}$0$}}\put(330,1229){\makebox(0,0)[r]{\strut{}$1$}}\put(787,220){\makebox(0,0){\strut{}C}}\put(1113,220){\makebox(0,0){\strut{}P}}\put(1438,220){\makebox(0,0){\strut{}O}}}\gplgaddtomacro\gplfronttext{\csname LTb\endcsname \put(1112,1937){\makebox(0,0){\strut{}Available Device }}\put(1112,1717){\makebox(0,0){\strut{} Information}}}\gplgaddtomacro\gplbacktext{\csname LTb\endcsname \put(2489,440){\makebox(0,0)[r]{\strut{}$0$}}\put(2489,1229){\makebox(0,0)[r]{\strut{}$1$}}\put(2963,220){\makebox(0,0){\strut{}L}}\put(3306,220){\makebox(0,0){\strut{}M}}\put(3648,220){\makebox(0,0){\strut{}H}}\put(3991,220){\makebox(0,0){\strut{}X}}}\gplgaddtomacro\gplfronttext{\csname LTb\endcsname \put(3477,1937){\makebox(0,0){\strut{}Attack Complexity }}\put(3477,1717){\makebox(0,0){\strut{}}}}\gplgaddtomacro\gplbacktext{\csname LTb\endcsname \put(5059,440){\makebox(0,0)[r]{\strut{}$0$}}\put(5059,1229){\makebox(0,0)[r]{\strut{}$1$}}\put(5533,220){\makebox(0,0){\strut{}N}}\put(5876,220){\makebox(0,0){\strut{}L}}\put(6218,220){\makebox(0,0){\strut{}M}}\put(6561,220){\makebox(0,0){\strut{}H}}}\gplgaddtomacro\gplfronttext{\csname LTb\endcsname \put(6047,1937){\makebox(0,0){\strut{}Supply Chain }}\put(6047,1717){\makebox(0,0){\strut{} Intrusion}}}\gplgaddtomacro\gplbacktext{\csname LTb\endcsname \put(7630,440){\makebox(0,0)[r]{\strut{}$0$}}\put(7630,1229){\makebox(0,0)[r]{\strut{}$1$}}\put(8116,220){\makebox(0,0){\strut{}IC}}\put(8470,220){\makebox(0,0){\strut{}II}}\put(8824,220){\makebox(0,0){\strut{}DC}}\put(9177,220){\makebox(0,0){\strut{}DI}}\put(9531,220){\makebox(0,0){\strut{}DA}}}\gplgaddtomacro\gplfronttext{\csname LTb\endcsname \put(8823,1937){\makebox(0,0){\strut{}Impact Attribute Weights}}\put(8823,1717){\makebox(0,0){\strut{} }}}\gplbacktext
    \put(0,0){\includegraphics[width={514.10bp},height={226.70bp}]{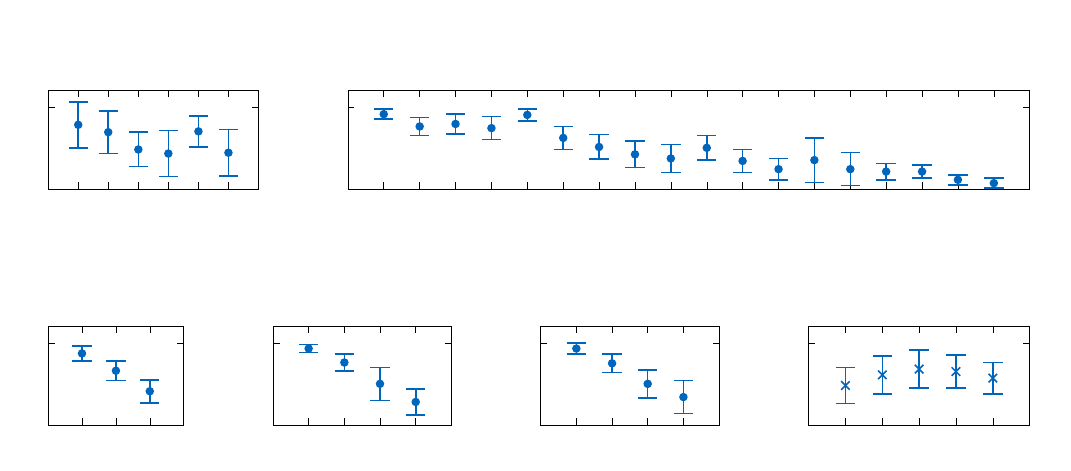}}\gplfronttext
  \end{picture}\endgroup
      \caption{Standard Deviation and mean {\color{blue}{$\bullet$}} / median ({\color{blue}{$\times$}}) of Numeric Values for Exploitability  and Attribute Weights for Impact}
    \label{fig:standarddev}
\end{figure*}

We define $V_\text{attr}$ as the numeric value associated with the \ac{CRESS} \cressattrval{Attr}{Value} in Table \ref{tab:CatValues}, whereas $w_\text{attr}$ is the attribute weight associated with the \ac{CRESS} \cressattr{Attr} in Table \ref{tab:CatValues}.
Weights not directly derived from interview results are indicated by an asterisk (*). 
We also show standard deviations of the \acp{DP} for the weights in Figure \ref{fig:standarddev}.

All weights and numeric values are derived by averaging over the available \acp{DP}.
For averaging, we used the mean or the median.
As the mean is not robust against outliers, it allows us to represent the disagreement of the interviewees' \acp{DP}. Instead, the median is robust against outliers and thus allows us to find consensus from the interview \acp{DP}.
For example, the \acp{DP} for end user \cressattrval{EP}{E} have a high spread because the interviewees' argumentation varied.
Most interviewees scored \cressattrval{EP}{E} as a highly critical entry point \cressattr{EP} because a malicious end user might be more common than, e.g. a malicious front end designer.
In contrast, some interviewees scored the end user \cressattrval{EP}{E} as an uncritical entry point \cressattr{EP}, as the end user's influence is limited to one or a few products. In contrast, a front end designer can compromise an entire batch of products.
This is also reflected in the high standard deviation for the Entry Point Attribute \cressattr{EP}, as shown in Figure \ref{fig:standarddev}. Other attributes, such as the RE Result \cressattr{RR} or Available Device Information \cressattr{AI} have a smaller standard deviation for their values, suggesting a similar view among interviewees regarding the impact of these values on the severity of the attack.
The arithmetic mean was used for averaging the \acp{DP} of the numeric values. Thus, we numerically represented the disagreement there.
However, the exploitability attribute weights \acp{DP} were averaged with the median to find a consensual compromise. As the impact numeric values were set empirically (see section \ref{ssec:equation_implementation}), the impact attribute weights were averaged using the mean so that interviewees' disagreement is not disregarded. Figure \ref{fig:standarddev} shows the mean and median for the exploitability numeric values and impact attribute weights.

The attribute weights $w_\text{attr}$ in Table \ref{tab:CatValues} show similar weights for the attributes entry point \cressattr{EP}, RE result \cressattr{RR}, attack complexity \cressattr{AC}, and supply chain intrusion \cressattr{SI}, while rating the attribute available device information \cressattr{AI} less severe.
For the possible and intended impacts ($\,$\cressattr{P**}, \cressattr{I**}$\,$), on-device Confidentiality \cressattr{*OC} was identified as having the greatest impact on severity, whereas attacks affecting IP Confidentiality \cressattr{*IC} were considered to be less critical.

Finally, to determine the final severity, we assessed the results for the overall weight of exploitability and impact.
We calculate the overall weight of exploitability as mean of the derived weights of the following attributes: entry point \cressattr{EP}, RE result \cressattr{RR}, available device information \cressattr{AI}, attack complexity \cressattr{AC}, and supply chain intrusion \cressattr{SI}.
The overall weight of impact could be derived directly from the interview \acp{DP}.
The results, as shown in Table \ref{tab:CatValues}, indicate that impact attributes are deemed to have a greater influence on attack severity compared to exploitability. This means that, according to our interviewees, the potential and intended impact ($\,$\cressattr{P**}, \cressattr{I**}$\,$) play a more crucial role in evaluating the seriousness of the attack.

We also derived the numeric values for all remediation strategy values.
However, the evaluation shows no significant difference when examining the means of the numeric values of the various types of remediation strategies, namely $0.41$ for pre-silicon remediation strategies \cressattrval{RS}{A**}, $0.48$ for detection-based post-silicon remediation strategies \cressattrval{RS}{*O*}, and $0.45$ for logic-based remediation strategies \cressattrval{RS}{**L}.
Also, deriving an individual numerical value for each remediation strategy separately, is not considered as a reasonable approach.
A very wide range of remediation strategies exist, and new methods are constantly developed, so any list would immediately be outdated or incomplete.
Thus, for our equation, we decided to consider only the fact of whether a remediation strategy exists for the attack or not, without considering graded numerical values for any concrete strategies.
This is also emphasized in Table \ref{tab:CatValues}.
We derive the numeric value for ``existing remediation strategy'' by taking the complement of the derived remediation strategy attribute weight: $1-0.37 = 0.63$.

\subsection{Equation Implementation}
\label{ssec:equation_implementation}

The interview results provide the basis for deriving the \ac{CRESS} scoring equation.
Based on the analysis in section \ref{ssec:equation_design_space}, we define the CRESS exploitability subscore 
\begin{align}
\begin{split}
\text{Exp.\,Score} = \big((V_\text{EP})^{w_\text{EP}} \cdot (V_\text{RR})^{w_\text{RR}} \cdot (V_\text{AI})^{w_\text{AI}}\\
    \cdot (V_\text{AC})^{w_\text{AC}} \cdot (V_\text{SI})^{w_\text{SI}}\big)^{\frac{1}{w_\text{EP}+w_\text{RR}+w_\text{AI}+w_\text{AC}+w_\text{SI}}}.
\end{split}
\label{eq:ExploitabilityScore}
\end{align}
Thus, the exploitability score is calculated as the weighted geometric mean of the exploitability attribute value weights. The multiplication ensures that the exploitability score is an intersection of the individual attributes, while the exponentiation ensures a weighted combination.

For the impact score, it is necessary to join the impact numeric value of the intended aim with the impact numeric value of the possible aim within each impact variant. To join the impacts, we define a impact sum
\begin{align}
    \overline{V}_\text{attr} = \min\big(V_\text{Intended Attr} + 0.5 \cdot V_\text{Possible Attr}, 0.9)\big)
\end{align}
Thus, the combined impact of intended and possible aim is a capped sum of the respective impact attribute numeric values, in which the possible aim will only contribute with half the intensity. This non-linear combination increases the influence of the possible impact iff the intended impact is not \cressval{H}.

With this, we define the impact subscore
\begin{align}
\begin{split}
    \text{Imp.\,Score} = 1-\bigg[
    \left(1-\overline{V}_\text{IC}\right)^{w_\text{IC}} 
    \cdot \left(1-\overline{V}_\text{II}\right)^{w_\text{II}}\\
    \cdot \left(1-\overline{V}_\text{OC}\right)^{w_\text{OC}}
    \cdot \left(1-\overline{V}_\text{OI}\right)^{w_\text{OI}}\\
    \cdot \left(1-\overline{V}_\text{OA}\right)^{w_\text{OA}}
    \bigg]^{\frac{1}{w_\text{IC}+w_\text{II}+w_\text{OC}+w_\text{OI}+w_\text{OA}}}.
\end{split}
\label{eq:ImpactScore}
\end{align}
Thus, the impact score is calculated as the complement of the weighted geometric mean of the complement of the capped impact sums. The calculation on complements produces a result similar to the union of the impact of the individual attributes, while the exponentiation weights the combination.

Finally, we define the total \ac{CRESS} score 
\newcommand{\sens}{\operatorname{sensitize}}\begin{align}
\label{eq:TotalCRESSSCore}
\text{CRESS\,Score} &=\\
10&\cdot\begin{split}
    \left\{\begin{array}{l}
    \big[0.65\cdot \sens(\text{Exp.\,Score})\cdot V_\text{RS}\\
    ~~+~0.84\cdot \text{Imp.\,Score}\cdot1.111\big] / 1.49,\\
    \hspace{2cm}\text{if Any Impact} \neq \cressval{N}\\
    0, \hspace{1.6cm}\text{else}
    \end{array}\right.
\end{split}\nonumber,
\end{align}
\begin{align}
\sens(x) &= \frac{1.0071}{1+\exp\left[-20(x-0.6372)\right]}
\end{align}
The total score is, in general, a weighted mean calculation on the exploitability and impact subscores, thus akin to a mixture distribution calculation. However, several adaptions were made to better match the aims stated in section \ref{ssec:equation_design_space}.

The $\sens$ transformation is a logistic function that ensures that the total score is sensitive to changes in exploitability. Only after this transformation the equation achieves the goal that a significant change in an exploitability attribute (eg. from \cressattrval{AC}{L} to \cressattrval{AC}{X}) results in a significant change in the total score (eg. from 6 to 5). The parameters of the logistic function were set empirically.

As the weighted geometric mean is 0 if one of its factors is 0, the complementation means that the impact subscore would be 1 if any impact is 1, so the equation would miss sensitivity to changes in any other impact attributes. To achieve sensitivity to changes in impact, we choose the numeric values for the impacts to be 0.9 at maximum. 
The other numeric values are chosen empirically, resulting in: \cressval{H}$= 0.90$, \cressval{L}$= 0.22$, and \cressval{N}$= 0.00$, see Table \ref{tab:CatValues}.
However, \cressval{H}$= 0.90$ means that the impact subscore can reach a value of 0.9 at maximum, so the total score would not achieve values in $[0,10]$. Thus, the scaling factor of $1.111$ ensures that the impact subscore is scaled to the interval $[0,1]$. Finally, the total score is scaled with $10$, so that it produces scores in the interval $[0,10]$. If exploitability and impact scores are displayed separately, they should also be multiplied by 10 for comparability.

\begin{figure}[!htbp]
     \centering
     \begingroup
  \makeatletter
  \providecommand\color[2][]{\GenericError{(gnuplot) \space\space\space\@spaces}{Package color not loaded in conjunction with
      terminal option `colourtext'}{See the gnuplot documentation for explanation.}{Either use 'blacktext' in gnuplot or load the package
      color.sty in LaTeX.}\renewcommand\color[2][]{}}\providecommand\includegraphics[2][]{\GenericError{(gnuplot) \space\space\space\@spaces}{Package graphicx or graphics not loaded}{See the gnuplot documentation for explanation.}{The gnuplot epslatex terminal needs graphicx.sty or graphics.sty.}\renewcommand\includegraphics[2][]{}}\providecommand\rotatebox[2]{#2}\@ifundefined{ifGPcolor}{\newif\ifGPcolor
    \GPcolortrue
  }{}\@ifundefined{ifGPblacktext}{\newif\ifGPblacktext
    \GPblacktexttrue
  }{}\let\gplgaddtomacro\g@addto@macro
\gdef\gplbacktext{}\gdef\gplfronttext{}\makeatother
  \ifGPblacktext
\def\colorrgb#1{}\def\colorgray#1{}\else
\ifGPcolor
      \def\colorrgb#1{\color[rgb]{#1}}\def\colorgray#1{\color[gray]{#1}}\expandafter\def\csname LTw\endcsname{\color{white}}\expandafter\def\csname LTb\endcsname{\color{black}}\expandafter\def\csname LTa\endcsname{\color{black}}\expandafter\def\csname LT0\endcsname{\color[rgb]{1,0,0}}\expandafter\def\csname LT1\endcsname{\color[rgb]{0,1,0}}\expandafter\def\csname LT2\endcsname{\color[rgb]{0,0,1}}\expandafter\def\csname LT3\endcsname{\color[rgb]{1,0,1}}\expandafter\def\csname LT4\endcsname{\color[rgb]{0,1,1}}\expandafter\def\csname LT5\endcsname{\color[rgb]{1,1,0}}\expandafter\def\csname LT6\endcsname{\color[rgb]{0,0,0}}\expandafter\def\csname LT7\endcsname{\color[rgb]{1,0.3,0}}\expandafter\def\csname LT8\endcsname{\color[rgb]{0.5,0.5,0.5}}\else
\def\colorrgb#1{\color{black}}\def\colorgray#1{\color[gray]{#1}}\expandafter\def\csname LTw\endcsname{\color{white}}\expandafter\def\csname LTb\endcsname{\color{black}}\expandafter\def\csname LTa\endcsname{\color{black}}\expandafter\def\csname LT0\endcsname{\color{black}}\expandafter\def\csname LT1\endcsname{\color{black}}\expandafter\def\csname LT2\endcsname{\color{black}}\expandafter\def\csname LT3\endcsname{\color{black}}\expandafter\def\csname LT4\endcsname{\color{black}}\expandafter\def\csname LT5\endcsname{\color{black}}\expandafter\def\csname LT6\endcsname{\color{black}}\expandafter\def\csname LT7\endcsname{\color{black}}\expandafter\def\csname LT8\endcsname{\color{black}}\fi
  \fi
    \setlength{\unitlength}{0.0500bp}\ifx\gptboxheight\undefined \newlength{\gptboxheight}\newlength{\gptboxwidth}\newsavebox{\gptboxtext}\fi \setlength{\fboxrule}{0.5pt}\setlength{\fboxsep}{1pt}\definecolor{tbcol}{rgb}{1,1,1}\begin{picture}(5016.00,6802.00)\gplgaddtomacro\gplbacktext{\csname LTb\endcsname \put(330,5321){\makebox(0,0){\strut{}$0$}}\put(759,5321){\makebox(0,0){\strut{}$1$}}\put(1188,5321){\makebox(0,0){\strut{}$2$}}\put(1617,5321){\makebox(0,0){\strut{}$3$}}\put(2046,5321){\makebox(0,0){\strut{}$4$}}\put(2475,5321){\makebox(0,0){\strut{}$5$}}\put(2903,5321){\makebox(0,0){\strut{}$6$}}\put(3332,5321){\makebox(0,0){\strut{}$7$}}\put(3761,5321){\makebox(0,0){\strut{}$8$}}\put(4190,5321){\makebox(0,0){\strut{}$9$}}\put(4619,5321){\makebox(0,0){\strut{}$10$}}}\gplgaddtomacro\gplfronttext{\csname LTb\endcsname \put(2474,6471){\makebox(0,0){\strut{}(a) CRESS Score for all combinations}}}\gplgaddtomacro\gplbacktext{\csname LTb\endcsname \put(330,3621){\makebox(0,0){\strut{}$0$}}\put(759,3621){\makebox(0,0){\strut{}$1$}}\put(1188,3621){\makebox(0,0){\strut{}$2$}}\put(1617,3621){\makebox(0,0){\strut{}$3$}}\put(2046,3621){\makebox(0,0){\strut{}$4$}}\put(2475,3621){\makebox(0,0){\strut{}$5$}}\put(2903,3621){\makebox(0,0){\strut{}$6$}}\put(3332,3621){\makebox(0,0){\strut{}$7$}}\put(3761,3621){\makebox(0,0){\strut{}$8$}}\put(4190,3621){\makebox(0,0){\strut{}$9$}}\put(4619,3621){\makebox(0,0){\strut{}$10$}}}\gplgaddtomacro\gplfronttext{\csname LTb\endcsname \put(2474,4771){\makebox(0,0){\strut{}(b) Exploit Score for all combinations}}}\gplgaddtomacro\gplbacktext{\csname LTb\endcsname \put(330,1920){\makebox(0,0){\strut{}$0$}}\put(759,1920){\makebox(0,0){\strut{}$1$}}\put(1188,1920){\makebox(0,0){\strut{}$2$}}\put(1617,1920){\makebox(0,0){\strut{}$3$}}\put(2046,1920){\makebox(0,0){\strut{}$4$}}\put(2475,1920){\makebox(0,0){\strut{}$5$}}\put(2903,1920){\makebox(0,0){\strut{}$6$}}\put(3332,1920){\makebox(0,0){\strut{}$7$}}\put(3761,1920){\makebox(0,0){\strut{}$8$}}\put(4190,1920){\makebox(0,0){\strut{}$9$}}\put(4619,1920){\makebox(0,0){\strut{}$10$}}}\gplgaddtomacro\gplfronttext{\csname LTb\endcsname \put(2474,3071){\makebox(0,0){\parbox{8.5cm}{\centering\strut{}(c) Exploit Score for all combinations without logistic function and remediation score}}}}\gplgaddtomacro\gplbacktext{\csname LTb\endcsname \put(330,220){\makebox(0,0){\strut{}$0$}}\put(759,220){\makebox(0,0){\strut{}$1$}}\put(1188,220){\makebox(0,0){\strut{}$2$}}\put(1617,220){\makebox(0,0){\strut{}$3$}}\put(2046,220){\makebox(0,0){\strut{}$4$}}\put(2475,220){\makebox(0,0){\strut{}$5$}}\put(2903,220){\makebox(0,0){\strut{}$6$}}\put(3332,220){\makebox(0,0){\strut{}$7$}}\put(3761,220){\makebox(0,0){\strut{}$8$}}\put(4190,220){\makebox(0,0){\strut{}$9$}}\put(4619,220){\makebox(0,0){\strut{}$10$}}}\gplgaddtomacro\gplfronttext{\csname LTb\endcsname \put(2474,1370){\makebox(0,0){\strut{}(d) Impact Score for all combinations}}}\gplbacktext
    \put(0,0){\includegraphics[width={250.80bp},height={340.10bp}]{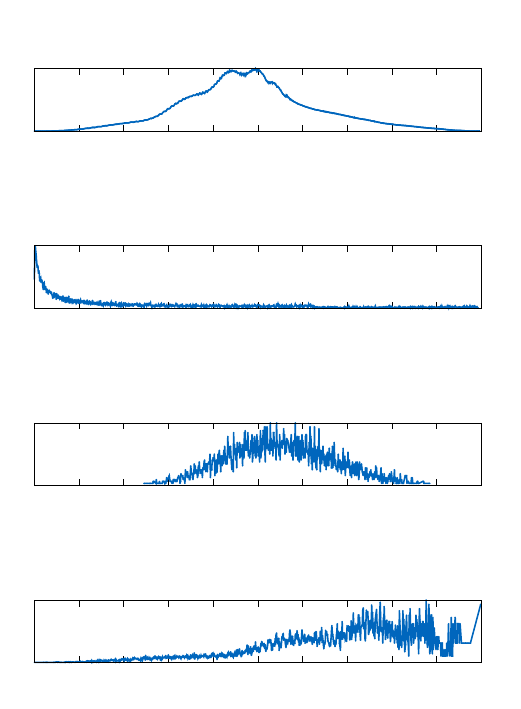}}\gplfronttext
  \end{picture}\endgroup
      \caption{Histogram of CRESS Scores for all combinations of attributes.}
    \label{fig:allValues}
\end{figure}

When the remediation strategy is treated as a binary decision, the total number of possible combinations across all attributes is 765,275,040. Figure \ref{fig:allValues} illustrates the analysis of the equation's outcomes for each of these combinations, using a histogram with 1,000 bins distributed evenly between 0 and 10. The distribution of the CRESS Score approximately follows a normal distribution centered around a mean value of approximately 5, as expected and desired from such a score. 
For the exploit score, we show both the final exploit score, which includes the remediation strategy and sensitization, as well as the original exploit score, without the remediation strategy and logistic function. We can observe the clear impact of sensitization. The final exploit score approximately follows a geometric distribution, while the impact variable approximately shows an inverted geometric distribution. Together, these result in the desired final score.

\section{Case Studies}
\label{sec:rating_case_studies}

In the following, selected case studies are rated. These are taken from the original CRESS paper \cite{Ludwig2021}, where they were described qualitatively. For a basic understanding, the studies are briefly explained and, subsequently, quantitatively discussed.

\subsection{RE-Improved Fault Injection}
\label{ssec:RE-improved_fault_injection}

\newcommand{\cressstring}[1]{
\par\noindent
\begingroup
\fbox{
\parbox{.95\linewidth}{
\sffamily\footnotesize
#1
}}
\endgroup\par\vspace{.5mm}\noindent
}

The first study is a RE-improved fault injection published by \textcite{Courbon2014}. In the exemplary scenario, a hardware-implemented AES cipher is attacked. For the attack, two devices are necessary. One is for reverse engineering of the device, and the other is for the subsequent laser fault injection. The reverse-engineered sample is prepared to the polysilicon layer, effectively exposing the structure of the gates. After scanning and two-dimensional stitching, flip-flops are detected through a pattern recognition algorithm. Via this approach, the location of the flip-flops enables a highly improved attack for which pre-localized areas are targeted. With additional knowledge about the AES implementation characteristics and timing information, the overall attack duration is reduced by a factor of  several hundred. This scenario is first qualitatively rated, allowing the deduction of the CRESS vector string:\\
\textbf{RE exploitability:~}In the scenario, the entry point is the end-user \cressattrval{EP}{E}. His/her target is the detection of standard cells \cressattrval{RR}{SI}, which is a relatively high target, building on intermediate targets such as the delayered, scanned, or stitched IC. Further, no additional information is available \cressattrval{AI}{C}.\\
\textbf{Attack exploitability:~}The laser fault injection requires specialized equipment, usually only found in dedicated laboratories \cressattrval{AC}{H}. The attack targets the end product, thus no supply chain intrusion is required \cressattrval{SI}{N}. There exists an effective defensive mechanism against this attack vector\cressattrval{RS}{A}. Possibilities are cell camouflaging to thwart the easy detection of flip-flops \cite{Gomez2019} or fault-resistant AES implementations \cite{Mestiri2019}.\\ 
\textbf{Impact.~}The main target of the attack is on-device confidentiality, aiming at the extraction of the secret AES key  \cressattrval{IOC}{H}. Further, RE reveals design information like the floorplan or information about the standard cell library, including the position of the attacked flip-flops resulting in \cressattrval{IIC}{H}. Further impact is unintended and minor caused as a by-product of the laser fault-injection \cressattrval{POI}{L}, \cressattrval{POA}{L}. This results in the following CRESS vector string:
\cressstring{EP:E/RR:SI/AI:C/AC:H/SI:N/RS:A/IIC:L/III:N/IOC:H/IOI:N/IOA:N/\\PIC:N/PII:N/POC:N/POI:L/POA:L/T:2023-12-01.}
The CRESS total score (Equation \ref{eq:TotalCRESSSCore}) of this scenario is 4.42, with an impact subscore (Equation \ref{eq:ImpactScore}) of 5.3 and an exploitability subscore (Equation \ref{eq:ExploitabilityScore}) of 6.6. 

\subsection{RISC-V Hardware Trojans}
\label{ssec:RISC-V_HT}

The next case study is a RISC-V implementation with hardware Trojans by \citeauthor{Hepp2021}~\cite{Hepp2021}. The authors taped out an open-source microcontroller design with post-quantum cryptographic accelerators. Four HTs infected the design, and the attacker is assumed to be a malicious IP provider of the microcontroller base design. For this example, we focus on one of the \acp{HT} that leaks arbitrary data through power or EM emanation. The rating of the attack is as follows:\\
\textbf{RE exploitability:~}The attacking party is the IP provider \cressattrval{EP}{I}. To be able to insert a functional HT, which enables the side-channel, the attacker requires the information from a partial functional identification (source of leaked data, trigger conditions, etc.) \cressattrval{RR}{PF}. The available device information in the scenario is open source \cressattrval{AI}{O}.\\
\textbf{Attack exploitability:~}Modifications in the supply chain are required by the addition of logic at the RTL level (read and write) and also for the corresponding software compiler framework (read and write), resulting in a high level of supply chain intrusion \cressattrval{SI}{H}. The side-channel can be exploited with equipment found in a standard, unspecialized laboratory \cressattrval{AC}{M}. Remediation strategies are available,e.g.~obfuscation or HT detection approaches \cressattrval{RS}{AO}.\\ 
\textbf{Impact.~}The HT insertion heavily affects two aspects. The first is the design integrity \cressattrval{III}{H} caused by the HT insertion. The second is on-device confidentiality caused by data leakage through the side-channel \cressattrval{IOC}{H}. Further minor, possible impacts are on-device integrity or availability \cressattrval{POI}{L}, \cressattrval{POA}{L} potentially caused by device overheating or voltage depletion caused by the HT execution. The attack vector yields following CRESS vector string:
\cressstring{EP:I/RR:PF/AI:O/AC:M/SI:H/RS:AO/IIC:N/III:H/IOC:H/IOI:N/IOA:N/\\PIC:N/PII:N/POC:N/POI:L/POA:L/T:2023-12-01.}
The CRESS total score of this scenario is 4.15, with an impact subscore of 7.4 and an exploitability subscore of 3.8. 

\subsection{Stealthy Dopant Hardware Trojans}
\label{ssec:stealthy_dopant_HTs}

The third case study also focuses on an HT insertion scenario titled stealthy dopant Trojans by \citeauthor{Becker2013}~\cite{Becker2013}. The HT scenario merely requires the modification of a single photomask or layer in the layout file -- the dopant mask or layer -- and subsequently enables covert power side channels. The strength of the attack is that the modification of dopant polarities, such as alterations of vertical interconnect access or metallization layers, cannot be detected via scanning electron microscopy. In the following, we focus on a specific attack scenario in the paper: the side-channel leakage of parametrically modified \emph{improved Masked Dual-Rail Logic (iMDPL)} gates. Through a change of dopant masks, a change of transistor areas leads to power side-channels. Via a correlation power analysis (CPA) on the SBox of an AES implementation, the key is leaked via this modification.\\
\textbf{RE exploitability:~}.The attacking party is a fab or foundry \cressattrval{EP}{F}. The attacking party must perform a partial functional identification \cressattrval{RR}{PF}. I.e., starting from the physical layout, the location and adjacent functionality of the targeted gates must be reverse-engineered for HT insertion. No available device information is necessary \cressattrval{AI}{C}.\\
\textbf{Attack exploitability:~}The attack complexity is high \cressattrval{AC}{H}. Specialized equipment is necessary to exploit the \emph{covert} side-channels. The supply chain intrusion is medium \cressattrval{SI}{M}, with read and write permission needed for the layout. Remediation strategies against this vector are available \cressattrval{RS}{AO}. Active strategies are e.g.~obfuscation techniques such as logic locking. 
An observing countermeasure is the passive voltage contrast of delayered devices \cite{Sugarawa2014}. 
\textbf{Impact.~}The impact is similar to the previous HT example. The major impact is the on-device confidentiality \cressattrval{IOC}{H} (side-channel leakage) and the IP integrity \cressattrval{III}{H} (layout tampering). A minor, possible impact is the on-device integrity and availability \cressattrval{POI}{L}, \cressattrval{POA}{L}.
The scenario results in the following CRESS vector string:
\cressstring{EP:F/RR:PF/AI:C/AC:H/SI:M/RS:AO/IIC:N/III:H/IOC:H/IOI:N/IOA:N/\\PIC:N/PII:N/POC:N/POI:L/POA:L/T:2023-12-01.}
The CRESS total score of this scenario is 4.16, with an impact subscore of 7.4 and an exploitability subscore of 4.0.

\begin{table*}[tbh]
\begin{minipage}{\linewidth}
\centering
\caption[Case Study Overview]{Overview of the five case studies with exploitability subscore, impact subscore, and total score. On the right, the scores are compared to CVSS v3.1\footnote{
The CVSS v3.1 exploitability was calculated as $^4\!\sqrt{AV\cdot AC\cdot PR\cdot UI}\cdot 10$ and the impact as $\left(1-^3\!\!\sqrt{(1-C)\cdot(1-I)\cdot(1-A)}\right)\cdot10$ to make these sub-scores' intervals comparable to those of CRESS.
}. The CVSS Medium rating interval is $4.0-6.9$.}
\label{tab:case_studies}
\begin{tabular}{@{}l|rrr|rrrrr@{}}
\toprule
\textbf{Case study}                          & \multicolumn{3}{c|}{\textbf{CRESS}} &\multicolumn{5}{c}{\textbf{CVSS v3.1}}\\
                          & Exploitability       & Impact & \textbf{Score} & Exploitability &Impact&Score&Rating&\\\midrule
\strut Fault attack \cite{Courbon2014}     & 6.6                  & 5.3  & \textbf{4.42} &5.02&3.56&5.3&Medium &\cite{CVSSFault}    \\
\strut RISC-V HTs \cite{Hepp2021}              & 3.8                  & 7.4  & \textbf{4.15}  & 3.77&3.56&4.9&Medium &\cite{CVSSRISCVHT}    \\
\strut Dopant HTs \cite{Becker2013}        & 4.0                  & 7.4  & \textbf{4.16}  & 3.77&3.56& 4.9&Medium &\cite{CVSSDopantHT}   \\
\strut IP theft                            & 5.4                  & 3.1  & \textbf{2.17}   &5.02&2.39& 4.9&Medium &\cite{CVSSIPTheft}  \\
\strut Vuln. detection                     & 5.3                  & 6.2  & \textbf{3.91}  &5.02&2.39&4.2&Medium &\cite{CVSSVulnDetect}\\
\strut \acs*{NVM} readout & 7.6 & 8.6 & \textbf{7.36} & 5.02&5.60&6.4&Medium& \cite{CVSSNVM} \\
\bottomrule
\end{tabular}
\end{minipage}
\end{table*}

\subsection{IP Infringement Scenarios}
\label{ssec:IP_infringement}

The next scenarios focus on IP infringement and have equal RE exploitability. The first scenario covers merely the theft of an exemplary proprietary design. For the second scenario, a vulnerability is detected and exploited via design analysis. The RE exploitability for both scenarios is the following: They are performed by an end user \cressattrval{EP}{E}. Partial functional identification \cressattrval{RR}{PF} of the specific module is the necessary identification target to enable either the theft and re-use of the IP or an attack on the IP. Both scenarios are carried on proprietary, closed source designs \cressattrval{AI}{C}.

\subsubsection{IP Theft of a Proprietary Multiplier Architecture}
\label{sssec:IP_theft}

This example does not include a subsequent attack, and the identification or \emph{theft} of the proprietary multiplier architecture is already the attack \cressattrval{AC}{N}, \cressattrval{SI}{N}. Remediation strategies exist \cressattrval{RS}{A} in the form of making the design extraction \emph{complex}. It can be achieved by obfuscation methods such as camouflaging, logic locking, or watermarking to prevent the piracy of proprietary IP. 
\\
\textbf{Impact.~}The only impact of this vector is IP confidentiality \cressattrval{IIC}{H}. All other categories are none. This results in the following CRESS vector string:
\cressstring{EP:E/RR:PF/AI:C/AC:N/SI:N/RS:A/IIC:H/III:N/IOC:N/IOI:N/IOA:N/\\PIC:N/PII:N/POC:N/POI:N/POA:N/T:2023-12-01.}
The CRESS total score of this scenario is 2.17, with an impact subscore of 3.1 and an exploitability subscore of 5.4.

\subsubsection{Vulnerability Detection in a Proprietary Cryptographic Algorithm}
\label{sssec:vulnerability_detection}

This case tackles vulnerability detection in a proprietary cryptographic algorithm. Via RE, the functionality of the algorithm is profoundly understood, enabling the vulnerability exploitation of it. We assume a trivial example: The design is identified to use the vulnerable SHA-1 hash function. For this, the attacker can use public collision attack strategies\cite{SBK+2017}.\\
\textbf{Attack exploitability:~}The subsequent attack may be executed with unspecialized equipment \cressattrval{AC}{L}. Further, no supply chain intrusion is needed \cressattrval{SI}{N}. Again remediation strategies exist \cressattrval{RS}{A}. They are the same as in the previous example (active), plus the usage of a secure algorithm can thwart an attack (e.g.~SHA-3).\\ 
\textbf{Impact.~}The impact of the attack is IP confidentiality \cressattrval{IIC}{H}, i.e.~the hash algorithm, and the on-device integrity \cressattrval{IOI}{H}, as the hashed data can be exchanged. This results in the following CRESS vector string:
\cressstring{EP:E/RR:PF/AI:C/AC:L/SI:N/RS:A/IIC:H/III:N/IOC:N/IOI:H/IOA:N/\\PIC:N/PII:N/POC:N/POI:N/POA:N/T:2023-12-01.}
The CRESS total score of this scenario is 3.91, with an impact subscore of 6.2 and an exploitability subscore of 5.3.

\subsection{Reverse Engineering Data in Non-Volatile Memory}
\label{sssec:re_otp}

The final case study considers an attack on a device with unprotected \ac{NVM}. Assume that the memory stores cryptographic keys both for on-device encryption and secure boot. The attack uses reverse engineering techniques to read the non-volatile data and aquire these keys (e.\,g.~\cite{CSW2017,QCF+2016}).

\textbf{RE exploitability:~}The attacking party is the end user \cressattrval{EP}{E}. The attacking party must aquire unstitched \ac{SEM} images of the \ac{NVM} to read the stored keys \cressattrval{RR}{US}. No available device information is necessary \cressattrval{AI}{C}.\\
\textbf{Attack exploitability:~}The attack complexity is medium \cressattrval{AC}{M}, \textcite{CSW2017} explicitly state that the equipment can be found in a standard, unspecialized laboratory. No supply chain intrusion is necessary \cressattrval{SI}{N}. Remediation strategies against this vector are available \cressattrval{RS}{A}. Active strategies are e.g.~obfuscation techniques or using tamper-proof memory such as with anti-fuses \cite{QCF+2016} or memory encryption.\\ 
\textbf{Impact.~}The impact both considers the readout of the device keys as a minor IP impact \cressattrval{IIC}{L}, as well as the total control over the device, as the security of encryption and secure boot is breached (\cressattrval{IO*}{H}).
The scenario results in the following CRESS vector string:
\cressstring{EP:E/RR:US/AI:C/AC:M/SI:N/RS:A/IIC:L/III:N/IOC:H/IOI:H/IOA:H/\\PIC:N/PII:N/POC:N/POI:N/POA:N/T:2023-12-11.}
The CRESS total score of this scenario is 7.36, with an impact subscore of 8.6 and an exploitability subscore of 7.6.

\subsection{Discussion}
\label{ssec:discussion}

The strength of CRESS is explicit when utilized comparatively. The comparison is illustrated in Table~\ref{tab:case_studies}. 

\subsubsection{CRESS renders attacks comparable}
The least-severe attack scenario is the IP-theft. While the exploitability is on a medium level, the low impact score leads to a low overall score. The most-severe attack is the readout of the NVM, because both exploitability as well as impact are severe.\\
Notably, the first three scenarios lead to similar overall scores. That means that by using CRESS, we can identify these attacks as similarly dangerous. On an absolute scale, the three attacks show medium danger. This is on-par with industry practice: The level of danger associated with such attacks is medium, because only at-risk systems are protected against fault and hardware trojan attacks.\\
The major difference between these scenarios becomes explicit when focusing on the exploitability and impact subscore. The attacks involving HTs (i.e.~RISC-V \cite{Hepp2021} and dopant HTs \cite{Becker2013}) have relatively high impact subscore. Yet, their low exploitability subscore indicates their exploitation requirements are \emph{high}. On the opposite, the Fault attack case study shows high exploitability, because the \ac{RE} effort is low, but the impact is more limited.

The IP-theft-based attack scenarios achieve low CRESS scores. While the exploitability yields a high score, the impact is limited. IP leakage of the mentioned proprietary multiplier module might be uncritical for end-users. Yet, the economic or reputation damage for the designer is significant. An attack that performs significant \ac{RE} to find the vulnerable use of SHA-1 bears little danger, as other means, such as a protocol analysis, can achieve this result without any \ac{RE}.

The last case study shows that even with moderate \ac{RE} effort, a high \ac{CRESS} score can result if the impact is significant. This is on par with industry best practice, as non-volatile memory for security-related purposes is always tamper-proofed.

\subsubsection{CRESS is more expressive than CVSS}

Using the CRESS score, the attack scenarios are much more comparable than the scores produced by CVSS. All case studies produce a rating in the medium range, three rate equally. The exploitabilities of all but the HT case studies are equal because the CVSS cannot represent the details of hardware attacks. The granularity of the impact is equally low. This is because the whole category of IP impact is not represented and the possibility of a further compromise using a non-patchable hardware exploit cannot be represented with CVSS. CVSS also gives obviously-wrong results because it rates the IP-theft case study more dangerous than the IP theft with an additional attack on the hardware---This is because in the IP-theft scenario, the IP-impact can only be represented by a scope change, else this scenario would be rated with zero impact.
We conclude that CVSS is not fit for rating hardware attack scenarios.

\subsubsection{CRESS is applicable to all hardware \ac*{CWE}} As the CRESS is more expressive than the CVSS, it provides advantages when evaluating hardware-based attack scenarios. The \ac{CWE} provides a broad taxonomy of weaknesses in hardware \cite{CWEHW}, ranging from \emph{manufacturing management} to \emph{physical access}. These listed weaknesses can, partially, be found using software-based discovery or random testing. These techniques do not involve \ac{RE}. Consequently, attack scenarios based on such attack vectors are out of the scope of CRESS.
However, all attack scenarios using any weakness listed in the Hardware Design CWE can be rated using CRESS, as soon as the process of weakness discovery uses any kind of hardware \ac{RE}. This renders CRESS broadly applicable to a wide range of attack scenarios, as most direct hardware attacks involve some kind of hardware \ac{RE}.

 \section{Conclusion}
\label{sec:conclusion}

We have enhanced the existing categorization of reverse engineering-based attacks by introducing a formula that assigns a score to describe the severity of these attacks. The formula was developed by consulting specialists to assign value to each \ac{CRESS} attribute and value, followed by a thorough evaluation of their responses. Higher-rated attacks are considered more severe, while lower-rated attacks are either difficult to exploit or have minimal impact. This scoring system not only allows us to categorize attacks based on their qualitative values in each category but also enables us to calculate a quantitative score for each type of attack. This is highly beneficial as it allows us to better compare different attack types and provides a quantifiable measure of the severity of new attacks. Additionally, it allows for the analysis of potential future attacks and the implementation of appropriate countermeasures. By leveraging the CRESS Score, we can develop accurate and tailored countermeasures for attacks that are deemed severe enough to warrant them. 
A web-based tool is available at \cite{CRESSCalculator}. The tool allows users to select CRESS attribute values according to their scenario to score it. The tool will then calculate the total CRESS score as well as the exploitability and impact subscores.

With the \ac{CRESS} Score, it is now possible to rate attacks based on weaknesses in the hardware \ac{CWE} database more expressively. CRESS is applicable as long as the attack does not fall outside the realm of reverse engineering and cannot be classified and quantified using the CRESS Score. Furthermore, while \ac{CRESS} provides a quantified, comparable, and useful rating, it is important to consider more granular ratings when necessary. Nevertheless, we believe that our approach covers a wide range of attacks that require some form of reverse engineering and provides a fair platform for rating them. Our methodology strikes a balance between generality for the purpose of comparison and specificity to allow for good differentiation. As with the \ac{CVSS}, we anticipate and hope for multiple iterations of the score as the community engages with and enhances it, fostering continuous improvement.

\ifCLASSOPTIONcompsoc
\section*{Acknowledgments}
\else
\section*{Acknowledgment}
\fi

We would like to sincerely thank our interviewees for their valuable contributions to this research: Nils Albartus, Navid Asadi Zanjani, Leonid Azriel, Steffen Becker, Ann-Christin Bette, Swarup Bhunia, Domenic Forte, Bernhard Lippmann, Samuel Pagliarini, Jofre Pallarès, Endres Puschner, Jonas Ruchti, Julian Speith, Zain Ul Abideen, René Walendy, and Edward Wang.
This work was partly funded by the German Ministry of Education and Research in the project VE-FIDES under Grant No.: 16ME0257.

\printbibliography[heading=References]

\end{document}